\documentclass[aps,amsmath,amssymb, notitlepage, 
twocolumn,
nofootinbib,
]{revtex4-1}

\usepackage{graphicx}
\usepackage{color}
\usepackage{bbold}

\newcommand{\bit}{\begin{itemize}}
\newcommand{\eit}{\end{itemize}}

\newcommand{\f}{\frac}
\renewcommand{\>}{\right\rangle}
\newcommand{\<}{\left\langle}
\newcommand{\ba}{\begin{align}}
\newcommand{\ea}{\end{align}}
\newcommand{\be}{\begin{equation}}
\newcommand{\ee}{\end{equation}}
\newcommand{\bi}{\begin{itemize}}
\newcommand{\ei}{\end{itemize}}
\newcommand{\lf}{\left(}
\newcommand{\ri}{\right)}

\newcommand{\CC}{\mathcal{C}}
\newcommand{\LL}{\mathcal{L}}

\newcommand{\FF}{\mathcal{F}}

\newcommand{\HH}{\mathcal{H}}

\newcommand{\ket}[1]{\left| #1 \>}
\newcommand{\bra}[1]{\< #1 \right|}

\newcommand{\up}{\uparrow}
\newcommand{\down}{\downarrow}

\newcommand{\T}{\mathcal{T}}
\newcommand{\SP}{\mathrm{Sp}}
\newcommand{\nn}{\\ \nonumber}

\begin{document}

\title{Topological Paramagnetism in Frustrated Spin-One Mott Insulators}

\author{Chong Wang, Adam Nahum, and T. Senthil}
\affiliation{Department of Physics, Massachusetts Institute of Technology, Cambridge, MA 02139, USA}
\date{\today}

\begin{abstract}
Time reversal protected three dimensional (3D) topological paramagnets are magnetic analogs of the celebrated 3D topological insulators.  Such paramagnets have a bulk gap, no exotic bulk excitations, but non-trivial surface states protected by symmetry. We propose that frustrated spin-$1$ quantum magnets  are a natural setting for realising such states in 3D. We describe a physical picture of the ground state wavefunction for such a spin-$1$ topological paramagnet in terms of loops of fluctuating Haldane chains with non-trivial linking phases. We illustrate some aspects of such loop gases with simple exactly solvable models. 
We also show how 3D topological paramagnets can be very naturally accessed within a slave particle description of a spin-$1$ magnet. Specifically we construct slave particle mean field states which are naturally driven into the topological paramagnet upon including fluctuations. We propose bulk projected wave functions for the topological paramagnet based on this slave particle description. An alternate slave particle construction leads to a stable $\mathrm{U}(1)$ quantum spin liquid from which a topological paramagnet may be accessed by condensing the emergent magnetic monopole excitation of the spin liquid.

\end{abstract}
\newcommand{\bea}{\begin{eqnarray}}
\newcommand{\eea}{\end{eqnarray}}
\newcommand{\p}{\partial}
\newcommand{\lp}{\left(}
\newcommand{\rp}{\right)}

\maketitle

Frustrated quantum magnets display a rich variety of many--body phenomena.  Some such magnets show long--range magnetic order at low temperature, often selected out of a  manifold of degenerate classical ground states by quantum fluctuations.  A very interesting alternative possibility --- known as quantum paramagnetism --- is the avoidance of such ordering even at zero temperature. Quantum paramagnets may be of various types. A fascinating and intensely--studied class is the quantum spin liquids: these display many novel phenomena, for instance fractionalization of quantum numbers and topological order, or gapless excitations that are robust despite the absence of broken symmetries \cite{sllee,slbalents,Wenbook}.  

Recently there has been much progress in understanding a different type of remarkable quantum paramagnet. These are phases which have a bulk gap and no fractional quantum numbers or topological order. Despite this, they have nontrivial surface states that are protected by global symmetries. These properties are reminiscent of the celebrated electronic topological band insulators. Hence they have been called topological paramagnets \cite{avts12}.  Topological paramagnets and topological band insulators are both examples of what are known as Symmetry Protected Topological (SPT) phases \cite{1dsptclass,1dz8,chencoho2011}. A classic example of a topological paramagnet is the Haldane/AKLT spin-$1$ chain: though this has a bulk gap and no bulk fractionalization, it has dangling spin-$1/2$ moments at the edge which are protected by symmetry, for instance time reversal.   
In the last few years tremendous progress has been made in understanding such SPT phases and their physical properties in diverse dimensions (for reviews, see Refs. \onlinecite{atav13,sptannrev}).

The main focus of the present paper is on three-dimensional topological paramagnets that are protected by time reversal (we also briefly  discuss topological paramagnets protected by other symmetries,  notably conservation of at least one spin component). These are interesting for a number of reasons. First, time reversal is a robust symmetry of typical physical spin Hamiltonians. In 1D the familiar Haldane/AKLT chain is the only time reversal protected topological paramagnet while in 2D there are no time reversal protected topological paramagnets. In 3D however there are three distinct non-trivial phases \cite{avts12,hmodl,burnellbc} (corresponding to a classification by the group $\mathbb{Z}_2^{\, 2}$). 
Second, 
regarded as an {\em electronic} insulator, unlike the 1D Haldane chain \cite{rosch}, these  3D topological paramagnets survive as distinct interacting SPT insulators \cite{wpssc14}. The properties  and experimental fingerprints of such topological paramagnets were described in Refs.~\cite{avts12,hmodl,burnellbc,wpssc14}.  However there is currently very little understanding  of where such phases might actually be found. In this paper we propose that frustrated spin-$1$ Mott insulators may be good places to look for an example of such phases. 

Already in the familiar 1D example it is  the spin-$1$ antiferromagnetic chain, rather than the spin-$1/2$ chain, that naturally becomes a topological paramagnet. In 3D for one of the topological paramagnets we provide a 
physical picture and a parton construction which are both very natural for the spin-$1$ case. We hope that our observations inspire experimental and numerical studies 
of frustrated spin-$1$ quantum magnetism in the future. Towards the end of the paper we remark on materials that may form such interesting frustrated magnets. 

The three 3D topological paramagnets that are protected by time reversal symmetry alone \cite{avts12, hmodl, burnellbc} all allow for a gapped surface with $\mathbb{Z}_2$ topological order ({\em i.e.} a gapped surface $\mathbb{Z}_2$ quantum spin liquid) even though the bulk itself is not topologically ordered.    The properties of this \emph{surface} theory give a useful way to label the {bulk} phases. The surface has gapped quasiparticle  excitations --- labelled  `$e$' and `$m$' ---  which are mutual semions. These may be thought of as the electric charge and magnetic flux of a deconfined $\mathbb{Z}_2$ gauge theory (like  the vertex and plaquette defects of Kitaev's toric code \cite{kitaev toric code}). At the SPT surfaces these particles have properties --- self-statistics or time reversal transformation properties --- that are impossible in a strictly 2D system, and which encode the topology of the bulk wavefunction.  The three nontrivial bulk states are denoted:
\ba \notag
&eTmT, &
& efTmfT,&
&efmf.
\end{align}
In the first and second, the surface $e$ and $m$ excitations are each Kramers doublets under time reversal, denoted by $T$. In the second and third they are fermions ($f$), while in the first they are bosons.  This paper focuses primarily on the `$eTmT$' state.

We begin by explaining a physical picture of a suitable ground state wave function for the $eTmT$ topological paramagnet. This is most easily visualized on a diamond lattice. We 
first close--pack each interpenetrating fcc sublattice of the diamond lattice with closed loops. On each loop we place all the spin-1 moments (located at the diamond sites) in the ground state of the 1D AKLT chain. We then superpose all such loop configurations with a crucial $(-1)$ sign factor whenever loops from the two different fcc sublattices link. We argue that this construction yields the topological paramagnet.

To understand the topological properties of such a wave function we describe a simple exactly solvable loop gas Hamiltonian \cite{lesiknote} --- equivalent to two coupled Ising gauge theories --- that clarifies the role of the `$(-1)^\text{linking}$' sign structure. In this solvable model the loops do not have AKLT cores but there are two species of loops on different sublattices with the mutual $(-1)$ linking sign. It demonstrates very simply how this sign leads to a state without intrinsic topological order. (This loop gas is not in  the $eTmT$ state, because of the absence of AKLT cores, but we show it to be  nontrivial in a different sense.)

Next we use the two--orbital fermionic parton representation developed for spin-$1$ magnets \cite{cenkeetal} to construct possible ground states.  When the fermionic partons have the mean--field dispersion of a certain topological superconductor, we show that the gauge fluctuations associated with the parton description convert the system into a topological paramagnet. In this construction the mean field state is {\em unstable} toward confinement by gauge fluctuations, as a result of a continuous nonabelian gauge symmetry. Despite this the bulk gap survives, leaving behind a non-trivial surface that we are able to identify as that of the $eTmT$ topological paramagnet. As a warm up exercise to illustrate some of the ideas of this 3D construction, we also describe how to access the 1D Haldane phase by confining a topological superconductor of parton fermions. The 3D construction naturally suggests alternative bulk wave functions for topological paramagnets, in the form of Gutzwiller--projected topological superconductors. This may be fruitful for future numerical work on the energetics of microscopic models. 

This parton construction also gives access to other SPT states for quantum magnets in 3D. For instance we show how to naturally obtain an SPT paramagnet  (dubbed $eCmT$ in Ref. \onlinecite{hmodl}) protected by  $\mathrm{U}(1) \times \mathbb{Z}_2^T$, where the $\mathrm{U}(1)$ describes rotation about one spin axis, say $S_z$, and $\mathbb{Z}_2^T$ is time reversal.  

Finally we show how to access a bulk $\mathrm{U}(1)$ quantum spin liquid with non-trivial implementation of time reversal symmetry.  Interestingly simply condensing the magnetic monopole of this $\mathrm{U}(1)$ spin liquid leads to an SPT state dubbed $eCTmT$ in the presence of both spin rotation and time reversal symmetries. If only time reversal is present this becomes the $eTmT$ state.

\section{Loop gas states}
\label{loop gas section}

In this section we describe a loop gas wavefunction that is naturally adapted to spin one magnets and  gives an intuitive picture for the $eTmT$ state. The wavefunction is a superposition of loop configurations, with each loop representing an AKLT state \cite{AKLT paper} for the spins lying on it. A given configuration enters the superposition with a sign factor determined by its topology: specifically, the loops come in two species, $A$ and $B$ (one associated with each sublattice of the bipartite diamond lattice) and the sign depends on the linking number of $A$ loops with $B$ loops. This geometrical picture makes the relationship between the bulk wavefunction and the surface excitations particularly simple. The surface $e$ and $m$ excitations are endpoints of the two species of AKLT chains, and are Kramers doublets since an AKLT chain has dangling spin-1/2s at its ends. 

In Sec.~\ref{pure loop section} we describe a similar wavefunction for `pure loops', i.e. loops that do not carry an internal AKLT structure. This may be regarded as a state of two coupled Ising gauge theories.
 It is \emph{not} in the $eTmT$ phase, but it illustrates the basic features of the loop gases in a simple model with an exactly solvable Hamiltonian. This `pure loop' model is also interesting in its own right:  when open strands (as opposed to closed loops) are banished from the Hilbert space, i.e. when  charge is  absent, it is in a nontrivial phase despite the absence of topological order. Therefore it may be viewed as a `constraint--protected' state. It would be interesting to relate this to the recent ideas of Ref.~\cite{kapustin thorngren higher symmetry}. We note that the models discussed in Ref.~\cite{burnell et al phase transitions} are also believed to be separated from the trivial phase by a phase transition, despite the absence of  topological order.

The wavefunctions discussed here are in a similar spirit to the Walker Wang models, which are formulated  in terms of string nets with a nontrivial sign structure, and show bulk confinement and surface topological order \cite{walker wang, keyserlingk et al surface anyons,burnellbc}. Constructions of SPTs using Walker Wang models were given in Refs.~\cite{burnellbc, chenanomaloussymm}. 2D `symmetry-enriched' topological states \cite{Yao Fu Qi, Huang et al SET, Li et al RAKLT} and SPT states \cite{avdomain} have also been constructed by attaching AKLT chains to loop-like degrees of freedom (see also \cite{wang xu two orbital boson}).

\subsection{Fluctuating AKLT chains}
\label{fluctuating AKLT chains}

\begin{figure}[b] 
\centering
\includegraphics[width=2.6in]{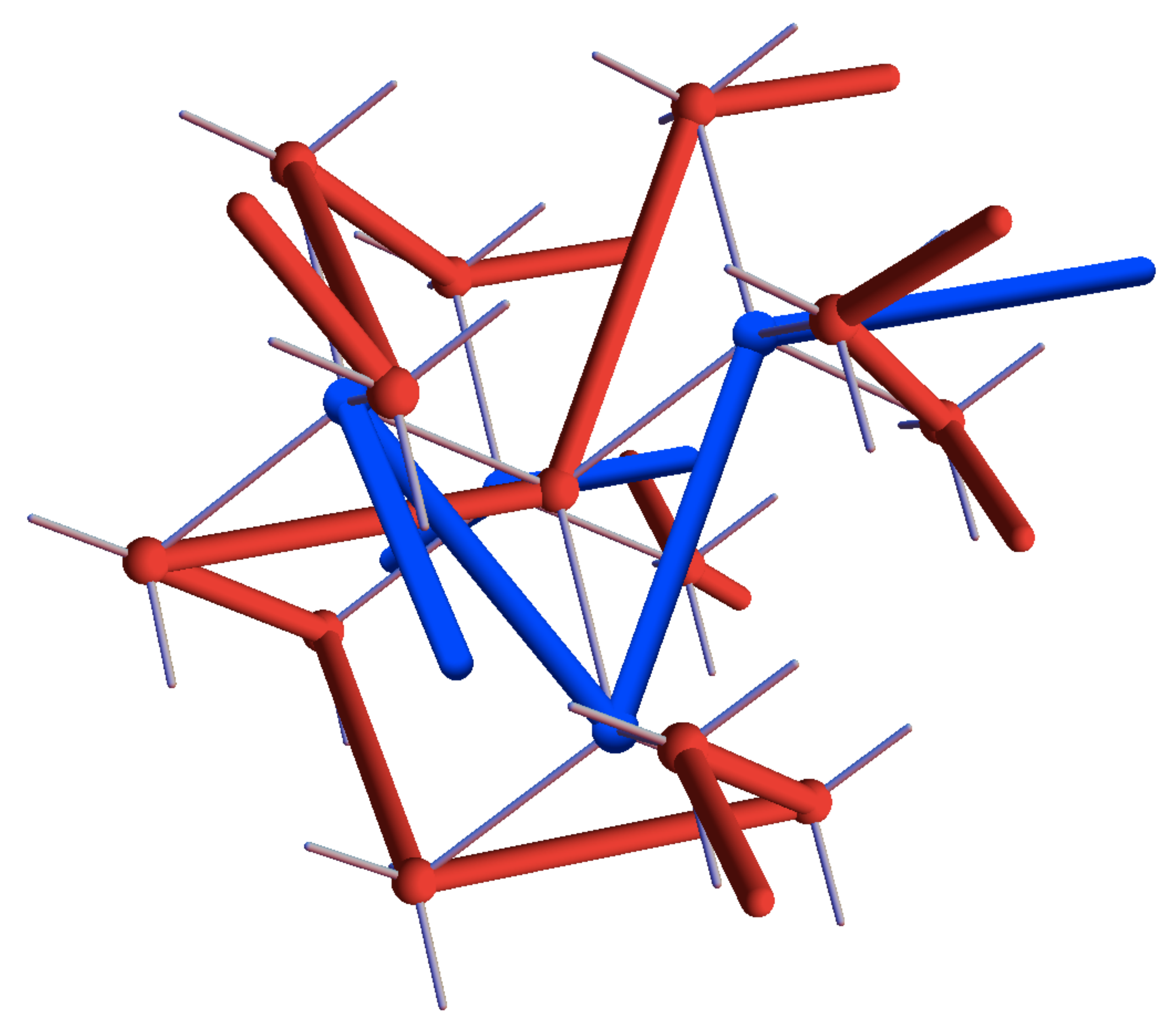}
\caption{Two species of AKLT loops, one on each sublattice of the diamond lattice. The loops live on the links of the fcc sublattices, i.e. on next-nearest-neighbour bonds of diamond.}
\label{diamondloops}
\end{figure}

The diamond lattice is made up of two fcc sublattices, $A$ and $B$.  If $\CC_A$ is a configuration of fully packed loops on $A$ (with every $A$ site visited by exactly one loop), we define $\ket{\CC_A}$ to be a product of AKLT states $\ket{\LL}$ for each of the loops $\LL$ in $\CC_A$,
\be
\ket{\CC_A} = \prod_{\mathcal{L} \in \CC_A} \ket{\LL}.
\ee
Similarly $\ket{\CC_B}$ is the state corresponding to a loop configuration $\CC_B$ on $B$. To define the AKLT states $\ket{\LL}$ fully we must choose an orientation for the fcc links, as discussed below (Sec.~\ref{Further details of AKLT state}).

Let $X(\CC_A, \CC_B)$ be the mutual linking number of the two species of loops. Since the loops are unoriented, this is defined modulo two: $X(\CC_A, \CC_B)=0,1$. A schematic wavefunction for the $eTmT$ phase may be written in terms of $X(\CC_A, \CC_B)$:
\ba\label{diamond wavefunction}
\ket{\Phi} &= 
\sum_{\CC_A, \CC_A} 
(-1)^{X(\CC_A, \CC_B)} \ket{\CC_A} \ket{\CC_B}.
\end{align}
For concreteness, we take periodic boundary conditions. The sums over $\CC_A$ and $\CC_B$ are then each restricted to loop configurations with an \emph{even} number of strands winding around the 3D torus in each direction, for reasons discussed below. This global constraint, together with the geometrical fact that the links of $A$ never intersect those of $B$, ensures that $X(\CC_A, \CC_B)$ is well defined.

The entanglement between the two sublattices in Eq.~\ref{diamond wavefunction} is entirely due to the sign factor. First consider what happens in the \emph{absence} of this sign factor. Each sublattice then hosts a superposition of loop configurations with positive amplitude, e.g. $\sum_{\CC_A} \ket{\CC_A}$. By analogy with the usual picture of deconfined $\mathbb{Z}_2$ gauge theory as a superposition of electric flux loop configurations \cite{fradkin book}, we would expect such a state to show $\mathbb{Z}_2$ topological order. (It is a 3D version of the `resonating AKLT' states studied in 2D \cite{Yao Fu Qi,Huang et al SET,Li et al RAKLT}.) The endpoint of an open AKLT chain is the deconfined $\mathbb{Z}_2$ charge in this state. Associated with the topological order is ground state degeneracy --- different ground states are distinguished by the parity of the winding number in each spatial direction.

In contrast, $\ket{\Phi}$ is \emph{not} expected to show topological order, despite the proliferation of long loops in Eq.~\ref{diamond wavefunction}. Instead it describes a phase in which the endpoints of open chains are confined in the bulk. Furthermore there is no ground state degeneracy: states with odd winding numbers are not ground states (i.e. are not locally indistinguishable from $\ket{\Phi}$).

More detailed discussion of this is deferred for the solvable model of Sec.~\ref{pure loop section}, but the basic idea is the following.  While the amplitude $(-1)^{X(\CC_A, \CC_B)}$ depends on the \emph{global} topology of the loop configurations, it amounts to the simple \emph{local} rule that the amplitude changes sign if an $A$ strand is passed through a $B$ strand. It is useful to imagine a hypothetical parent Hamiltonian that imposes this sign rule. But the sign rule cannot be consistently imposed if the wavefunction includes open strands or configurations with odd winding numbers (see below). Similar phenomena occur in the confined Walker--Wang models \cite{walker 
wang, keyserlingk et al surface anyons, burnellbc}.

\begin{figure}[t] 
\centering
\includegraphics[width=3.3in]{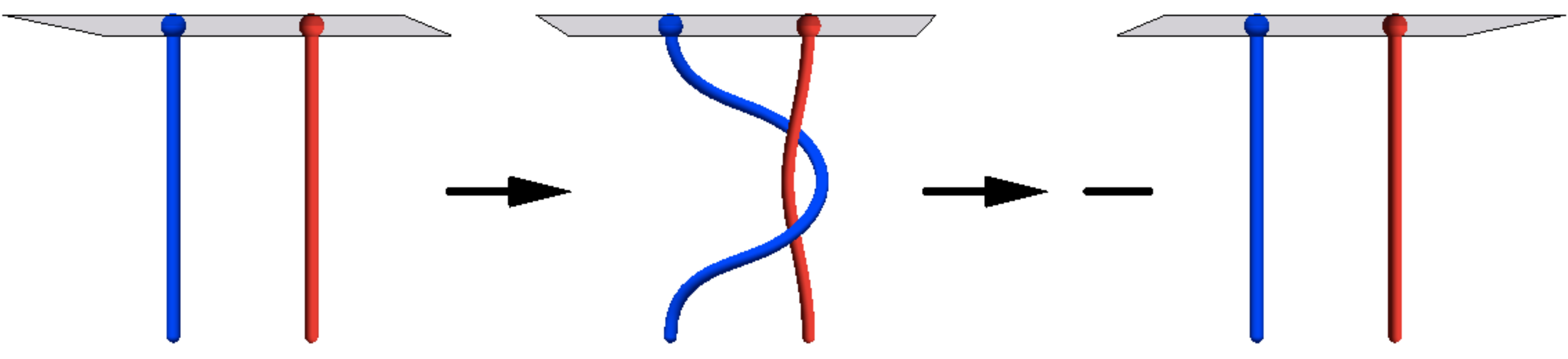}
\caption{For appropriate boundary conditions, endpoints of $A$ and $B$ chains (red and blue respectively) give surface excitations with mutual semionic statistics. Braiding the anyons on the surface (first arrow) changes the sign of the wavefunction, for consistency with the rule that configurations related by passing an $A$ strand through a $B$ strand in the bulk (second arrow) appear in the wavefunction with opposite sign.}
\label{surface braiding}
\end{figure}

However, open endpoints are deconfined at the boundary, for appropriate boundary conditions. The minus sign associated with passing an $A$ strand through a $B$ strand in the bulk means that the endpoints are mutual semions \cite{xuts13} --- see Fig.~\ref{surface braiding}. They are also Kramers doublets. These surface properties are the defining features of the $eTmT$ state. The  wavefunction $\ket{\Phi}$ has more symmetry than simply time reversal (e.g. separate spin rotation symmetries for each sublattice) but if it is indeed in the $eTmT$ phase then these symmetries could be weakly broken without leaving the phase.

\subsection{Further details on fluctuating AKLT state}
\label{Further details of AKLT state}

To write the AKLT-based state explicitly it is convenient to represent the spin-one at each site $i$ in terms of auxiliary spin-1/2 bosons \cite{AKLT paper, wang xu two orbital boson}. If the boson creation operators are $b^\dag_{i\alpha}$ ($\alpha = \uparrow, \downarrow$), then $\vec S_i = \f{1}{2} b^\dag_{i\alpha} \vec \sigma_{\alpha\beta} b_{i\beta}$. The occupation number $b_{\alpha i}^\dag b_{i\alpha}$ is equal to two to   ensure spin one at each site. The AKLT state $\ket{\LL}$ is then created by acting on the boson vacuum with operators $S_{ij}^\dag$ that create singlet pairs on the links of the loop, which we normalize  as $S^\dag_{ij}=  \f{1}{\sqrt 3} ( b^\dag_{i\uparrow} b^\dag_{j\downarrow} - b^\dag_{i\downarrow} b^\dag_{j\uparrow})$.  This operator is antisymmetric in $(i,j)$, so to define $\ket{\Phi}$ we must fix an orientation for the links of each fcc sublattice. (The fcc lattice has four sublattices, $a$, $b$, $c$, $d$, so for example we could orient the links from $a\rightarrow b$, $a\rightarrow c$, $a\rightarrow d$, $b\rightarrow c\rightarrow d\rightarrow b$, with the orientations on each sublattice related by inversion symmetry.) Then for each sublattice
\be
\ket{\CC} =  \prod_{\<ij\> \in \CC} S^\dag_{ij} \ket{\text{vac}},
\ee
where $i$ is the site at the tail of the oriented link $\<ij\>$. These states satisfy $\langle \CC | \CC \rangle =  \prod_\text{loops} ( 1 + {(-1)^\ell}/{3^{\ell -1}} )$, where $\ell$ is the length of a given loop \cite{AKLT paper}.

It should be noted that that expectation values in the state $\ket{\Phi}$ are nontrivial, in particular because overlaps $\langle \CC | \CC' \rangle$ for distinct $\CC$, $\CC'$ are nonzero. So while it is plausible that $\ket{\Phi}$ is in the $eTmT$ phase, this cannot be established purely analytically. For example, the state could in principle break spatial or spin rotation symmetry spontaneously. A cautionary example is given by the uniform-amplitude resonating valence bond state for spin-1/2s on the cubic lattice: this has weak N\'eel order \cite{cubic RVB numerics}, despite being a superposition of singlet configurations which individually have trivial spin correlations. In the present model, the entanglement between sublattices  supresses off-diagonal elements of the reduced density matrix when written in the AKLT-chain basis \cite{reduced density matrix note}. Together with the non-bipartiteness of the fcc lattice, this makes spin order seem less likely. But since $\ket{\Phi}$ is intended to illustrate 
the topological structure of the phase, and not as a ground state of a realistic Hamiltonian, it may not be crucial whether it is in the desired phase as written or whether further tuning of the amplitudes is required.

\section{`Pure loop' state}
\label{pure loop section}

\begin{figure}[t] 
\centering
\raisebox{0.0\height}{
\includegraphics[width=1.5in]{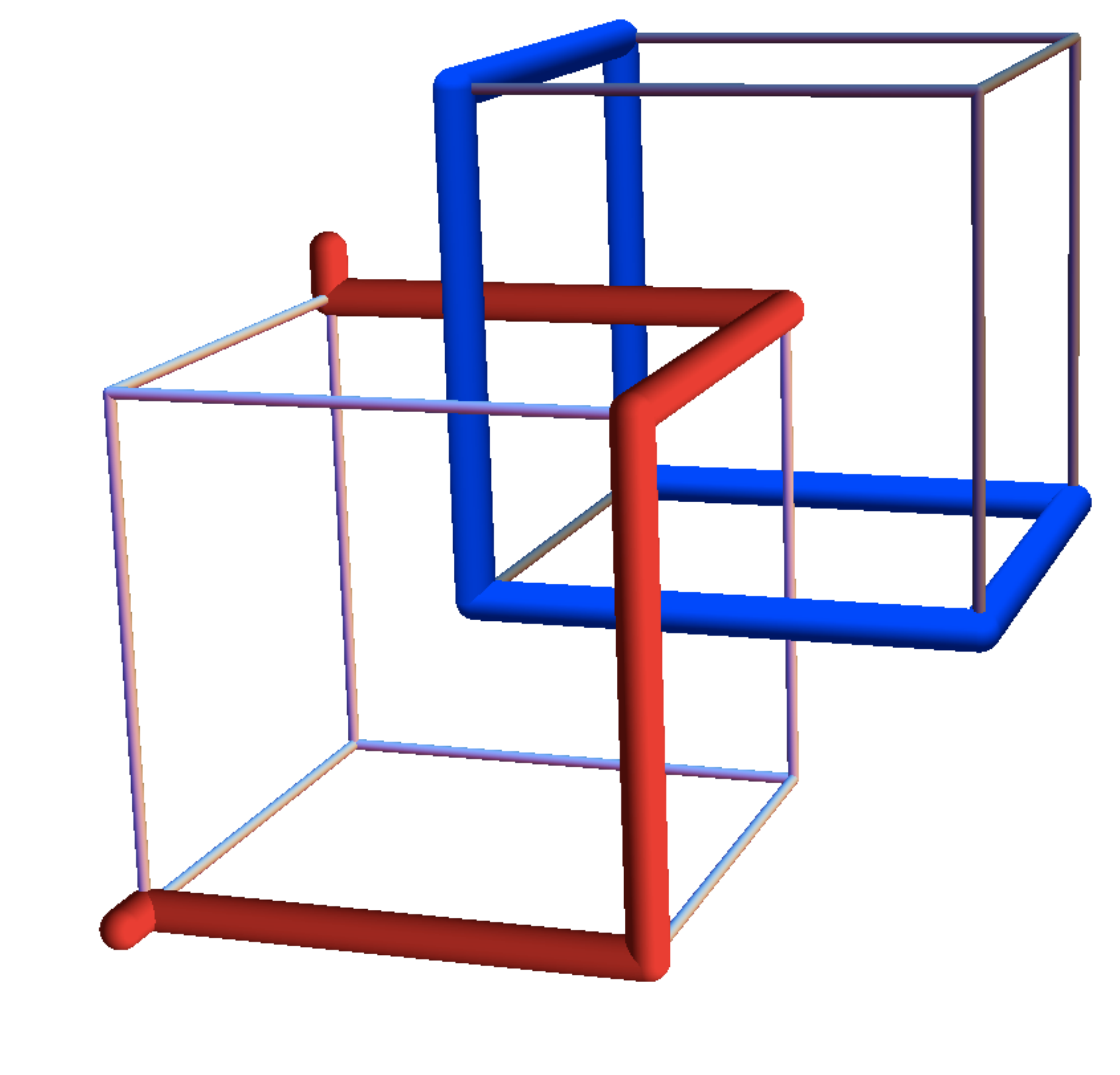}
}
\raisebox{0.4\height}{
\includegraphics[width=1.4in]{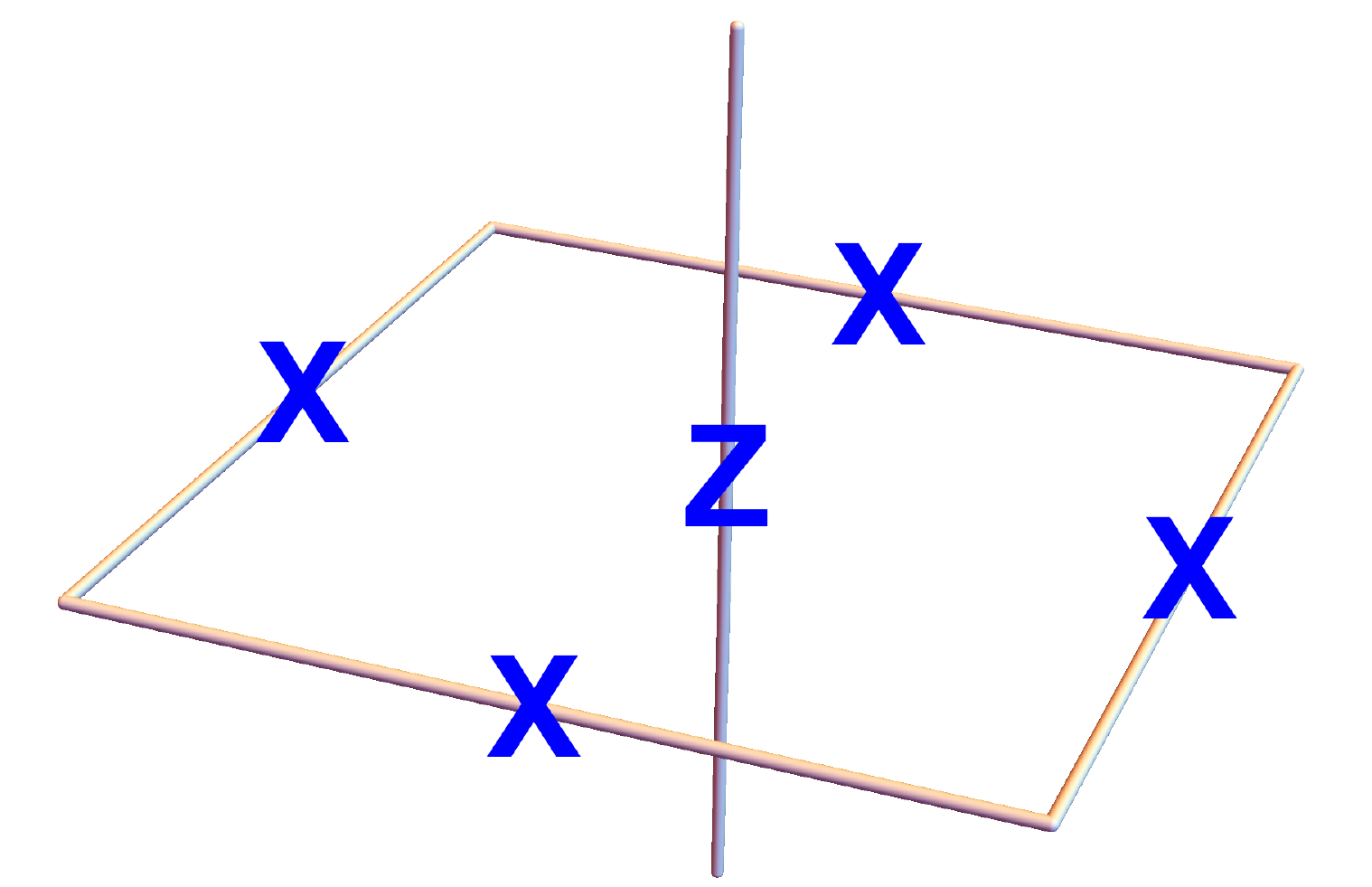}
}
\caption{Left: Loops on interpenetrating cubic lattices $A$ and $B$. The state $\ket{\Psi}$ is a superposition of such configurations with signs determined by linking of $A$ and $B$ loops. Right: the product of Pauli matrices defining the flip term $\FF$ on a plaquette (see Eq.~\ref{flip operators}).}
\label{twocubes}
\end{figure}

It is enlightening to look at the simplest model\cite{lesiknote} that captures the $(-1)^\text{linking}$ sign structure. To this end we take a system of spin-1/2s on the \emph{links} of two interpenetrating cubic lattices $A$ and $B$, as shown in Fig.~\ref{twocubes}.  We think of a down spin (in the `$z$' basis) as an occupied link, and an up spin as an unoccupied one. The number of occupied links at each vertex is always even in the state we consider, so the configurations of occupied links,  $\CC_A$ and $\CC_B$, can be decomposed into closed loops.\footnote{With a harmless ambiguity when the number of occupied links at a vertex exceeds two.} We refer to $\CC_A$ and $\CC_B$ as loop configurations.  Other solvable loop gas/string net models have been considered in Refs.~\cite{keyserlingk et al surface anyons,burnellbc}, using the Walker Wang construction \cite{walker wang}.

The `pure loop' state analogous to $\ket{\Phi}$ above is (again we sum only over loop configurations with even winding numbers on each sublattice):
\be\label{pure loop state}
\ket{\Psi} = \sum_{\CC_A, \CC_B} (-1)^{X(\CC_A,\CC_B)} \ket{\CC_A} \ket{\CC_B}.
\ee
We may view $\CC_A$ and $\CC_B$ as the electric flux line configurations for a pair of coupled $\mathbb{Z}_2$ gauge fields, with one $\mathbb{Z}_2$ gauge field living  on each cubic lattice. Imposing the above sign structure for the two sets of electric flux lines is equivalent to binding the  electric flux line of each gauge field to the magnetic flux line of the other, as will be clear shortly.

It is straightforward to write down a gapped parent Hamiltonian $\HH_\text{linking}$ for $\ket{\Psi}$,  using the fact that  flipping the occupancy of all the links on the plaquette changes the linking number $X(\CC_A,\CC_B)$ if and only if the link piercing the plaquette is occupied.  $\HH_\text{linking}$ is a sum of terms for the plaquettes $p$ of each cubic lattice:
\be \label{plaquette hamiltonian}
\HH_\text{linking} = -\left(
J \,  \sum_{p \in A}  \, \FF_{Ap}
+ J \,  \sum_{p \in B}  \, \FF_{Bp}\right).
\ee
The operators $\FF_A$ and $\FF_B$ flip the occupancy of the links on a plaquette, with a sign that depends on whether the link piercing it is occupied. Allowing $p$ to denote both a plaquette and the link piercing it, and denoting the Pauli operators on $A$ and $B$ by $\vec \sigma$ and $\vec \tau$ respectively,
\ba\label{flip operators}
\FF_{Ap} & = \tau^z_p \,\,  \prod_{l\in p} \sigma^x_l, &
\FF_{Bp} & = \sigma^z_p \,\,  \prod_{l\in p} \tau^x_l.
\end{align}
These operators all commute, so the Hamiltonian is trivially solvable. $\ket{\Psi}$ is the unique ground state and minimises each term of $\HH_\text{linking}$ since $\FF \ket{\Psi} = \ket{\Psi}$ for each plaquette operator. 

The state $\ket{\Psi}$ contains only closed loops, i.e. it satisfies
\ba\label{vertex constraints}
\prod_{l\in v} \sigma^z_l &=1& &\text{for $v\in A$}, &  &&
\prod_{l\in v} \tau^z_l &=1& &\text{for $v\in B$}
\end{align}
where $v$ denotes a vertex and $l \in v$ the links touching $v$. Any state  satisfying $\FF\ket{\Psi}=\ket{\Psi}$ for all the plaquette operators must also satisfy these vertex conditions, because $\prod_{l\in v} \sigma^z_l$ and $\prod_{l\in v} \tau^z_l$ can be written as products of $\FF$s.

We may regard Eqs.~\ref{vertex constraints} as the gauge constraints for a pair of pure $\mathbb{Z}_2$ gauge theories (the $\mathbb{Z}_2$ versions of $\vec \nabla.\vec E=0$). The two electric fields are given by $\sigma^z$ and $\tau^z$ and live on the links of $A$ and $B$ respectively. The {magnetic} field of each gauge field lives on the  links of the \emph{opposite} lattice to its electric field. For example the magnetic field of $\sigma$ is given by $\prod_{l\in p}\sigma^x$, where $p$ is a plaquette of $A$, or equivalently a link of $B$.

In this language, $\HH_\text{linking}$ simply glues the electric flux line of each species to the magnetic flux line of the other. The $\sigma$--magnetic flux and the $\tau$--electric flux are equal since $\FF_A=1$, and the  $\sigma$--electric  and  $\tau$--magnetic fluxes are equal via $\FF_B=1$.

The state $\ket{\Psi}$ is not topologically ordered. Neither is it a time-reversal protected SPT: it can be adiabatically transformed to a product state without breaking time reversal symmetry. However it \emph{is} protected if impose Eqs.~\ref{vertex constraints} as constraints: i.e. if we forbid open strands, as opposed to closed loops. In  the gauge theory language, this means forbidding charge. With this constraint it is impossible to reach a trivial state without going through a phase transition, as follows from the self--duality of the state described in Sec.~\ref{self-duality}.

We will explain these features from several points of view below. 
One convenient approach which leads to a geometric picture is to switch from the $(\sigma^z, \tau^z)$ basis used in Eq.~\ref{pure loop state} to the $(\sigma^z, \tau^x)$ basis. The $\sigma^z$ configuration is a loop configuration on the $A$ lattice, as above. We represent the $\tau^x$ configuration by a configuration of \emph{membranes} made up of plaquettes on the $A$ lattice: $\tau^x_p=-1$ represents an occupied plaquette, and $\tau^x_p = 1$  an unoccupied one. 

The $\FF_{B}$ terms in $\HH_\text{linking}$ act on a link of the $A$ lattice together with the four plaquettes touching it. $\FF_{B}=1$ imposes the rule that the $\sigma^z$ loops are glued to the boundaries of the $\tau^x$ membranes, i.e. to the links where an odd number of occupied plaquettes meet. This is  the gluing of $\sigma$--electric flux lines (where $\sigma^z=-1$) to $\tau$--magnetic flux lines (where $\prod \tau^x = -1$) mentioned above.

\begin{figure}[t] 
\centering
\includegraphics[width=2.1in]{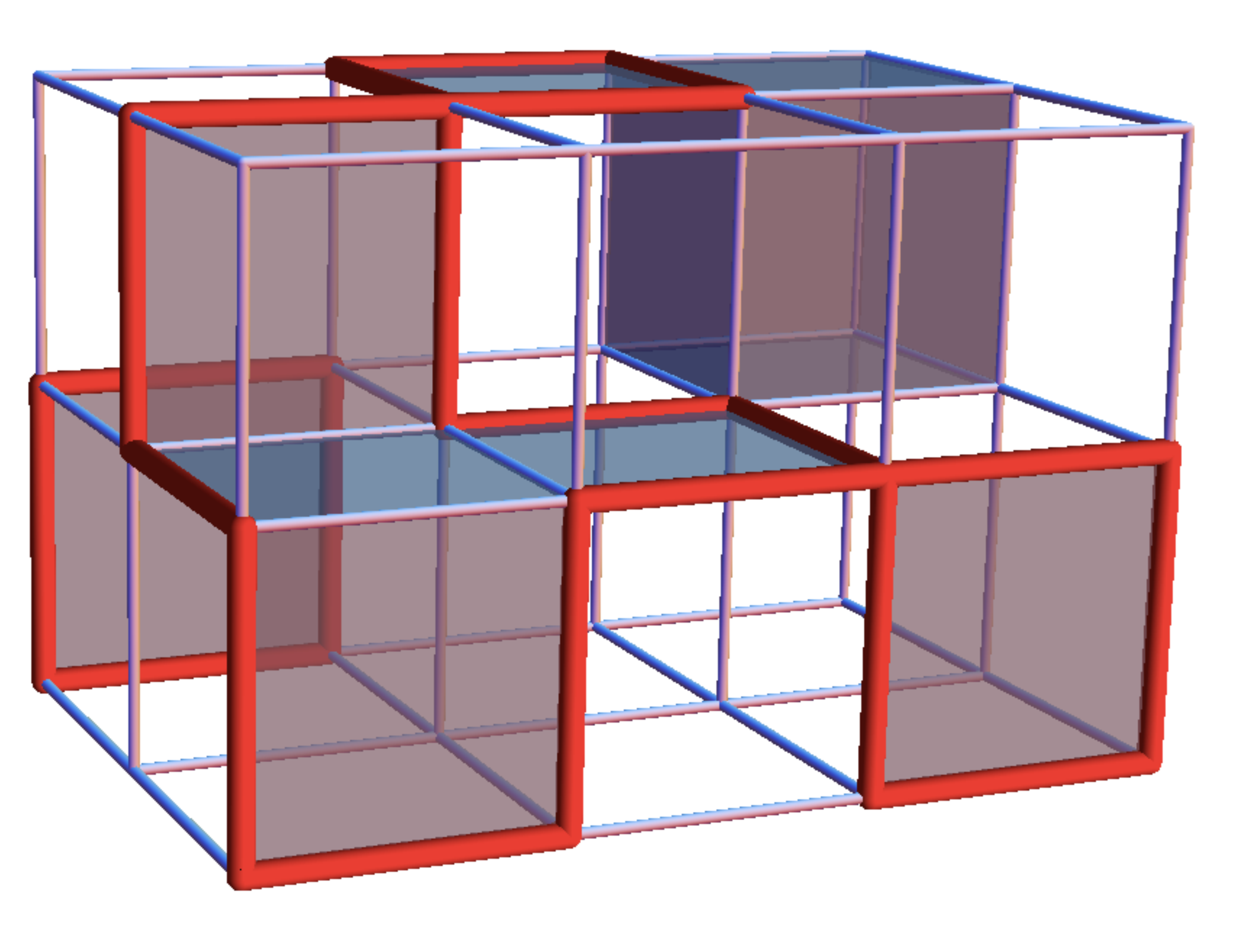}
\caption{After a basis change, $\ket{\Psi}$ is a superposition of membrane configurations ($\tau^x = -1$ on shaded plaquettes) with red loops (where $\sigma^z = -1$) glued to membrane boundaries. (The red loops are $\sigma$--electric lines and the membrane boundaries are $\tau$--magnetic lines.)}
\label{surfaces}
\end{figure}

Let $\mathcal{M}$ denote a membrane configuration, and  $\ket{\mathcal{M}}$  the corresponding state with $\protect{\tau^x = -1}$ on the occupied plaquettes. Let $\partial \mathcal{M}$ be the loop configuration given by the boundaries of the membranes in $\mathcal{M}$. Then $\ket{\Psi}$ can be written  (neglecting an overall constant)
\be\label{surface representation of pure loop state}
\ket{\Psi} = \sum_\mathcal{\CC_A} 
\sum_{\substack{
\mathcal{M} \\
\partial \mathcal{M} = \CC_A }}
 \ket{\CC_A}
\ket{\mathcal{M}}.
\ee
Fig.~\ref{surfaces} shows the geometrical interpretation of this state. It is a soup of $\tau^x$ membranes, with $\sigma^z$ loops glued to their boundaries.

Confinement of string endpoints is easy to see in this basis. A pair of vertex excitations at which $\prod_{l\in v} \sigma_l^z = -1$ are connected by an open string.  Since the boundary of $\mathcal{M}$ contains only closed loops, the open string makes it impossible to satisfy the gluing of strings to membrane boundaries demanded by the $\FF_B$ terms in $\HH_\text{linking}$. If the separation of the vertex defects is $D$, there must be at least $D$ unsatisfied links, giving a linear confining potential for such defects. For similar reasons, a configuration with an odd number of winding $\sigma^z$ strands in some direction costs an energy proportional to the spatial extent of the system in this direction. By symmetry, this applies equally to the $\tau^z$ strings that are present in the original basis. 

We can also understand the confinement of string endpoints algebraically (Refs.~\cite{keyserlingk et al surface anyons,burnellbc} give analogous arguments for bulk confinement and surface topological order in the Walker Wang models).  The Hamiltonian in Eq.~\ref{plaquette hamiltonian} is clearly exactly soluble not just for the ground state but for all excited states. An `elementary' excitation is given by a `defect' in some square plaquette, say on the B lattice, with 
\begin{equation}
\FF_{Bp} = -1
\end{equation}
while $\FF= +1$ on all other plaquettes of either sublattice. Such a defect plaquette costs energy $2J$.  It leads to a violation of the closed loop vertex constraint for $\sigma^z$ on the two vertices of the $A$ sublattice connected by the $A$-link that penetrates the defect plaquette. Thus the excitation we have created has two string end-points on nearest neighbor A-sites. To move these string endpoints apart by a distance $D$ we must create $o(D)$ such defect plaquettes. Consequently the energy cost is also $o(D)$ and we have linear confinement of string endpoints.

In the gauge theory language, the reason for the absence of deconfined excitations is that the tensionless lines in this state are not lines of  pure electric flux, but rather  of electric flux together with magnetic flux of the other species. If such lines could end, their endpoints would be deconfined excitations. But the Hilbert space does not allow for such excitations: a magnetic flux line cannot terminate in the bulk (by virtue of its definition in terms of e.g. $\prod \tau^x$).    
 
\begin{figure}[b] 
\centering
\includegraphics[width=1.65in]{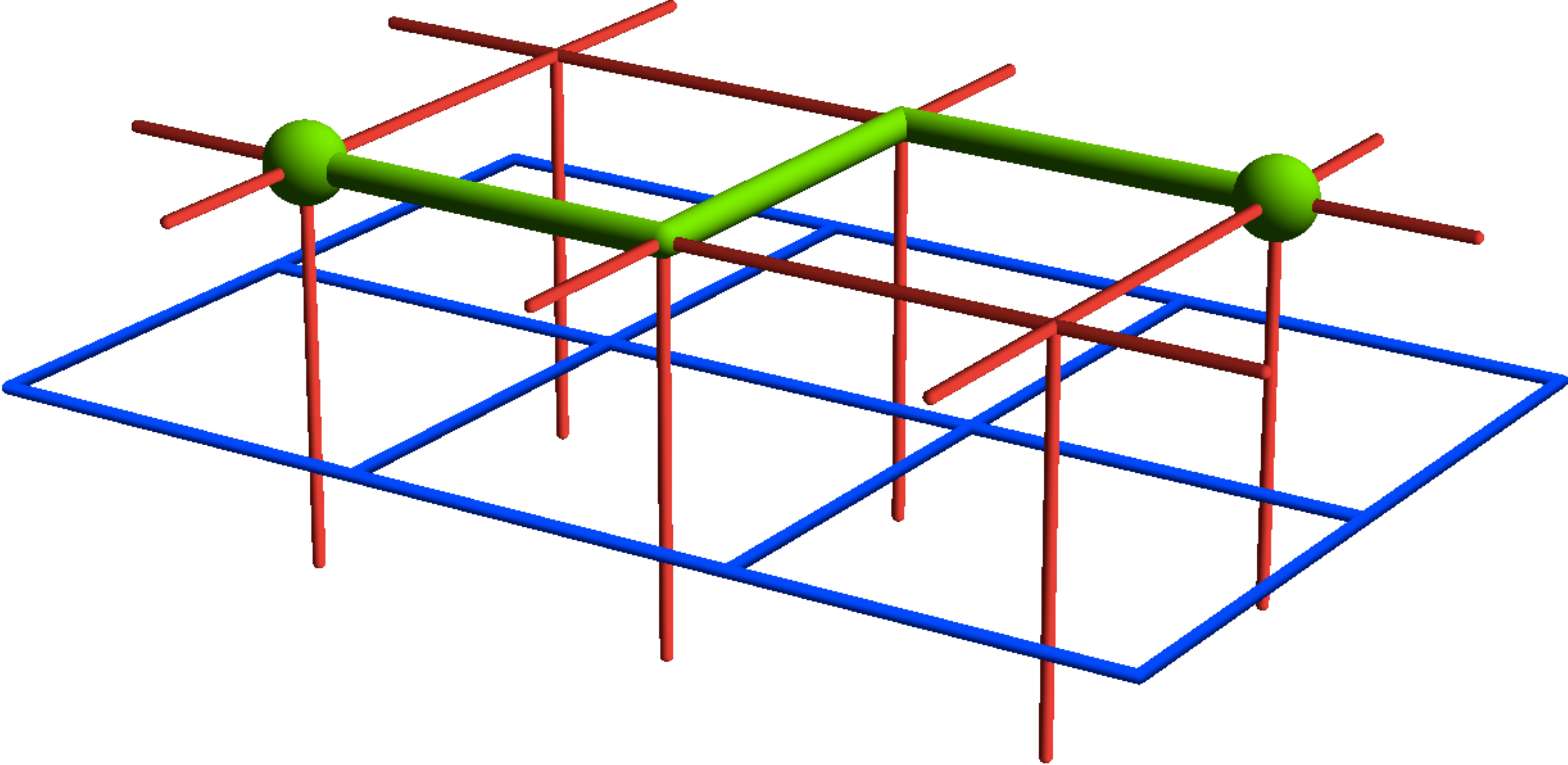}
\includegraphics[width=1.65in]{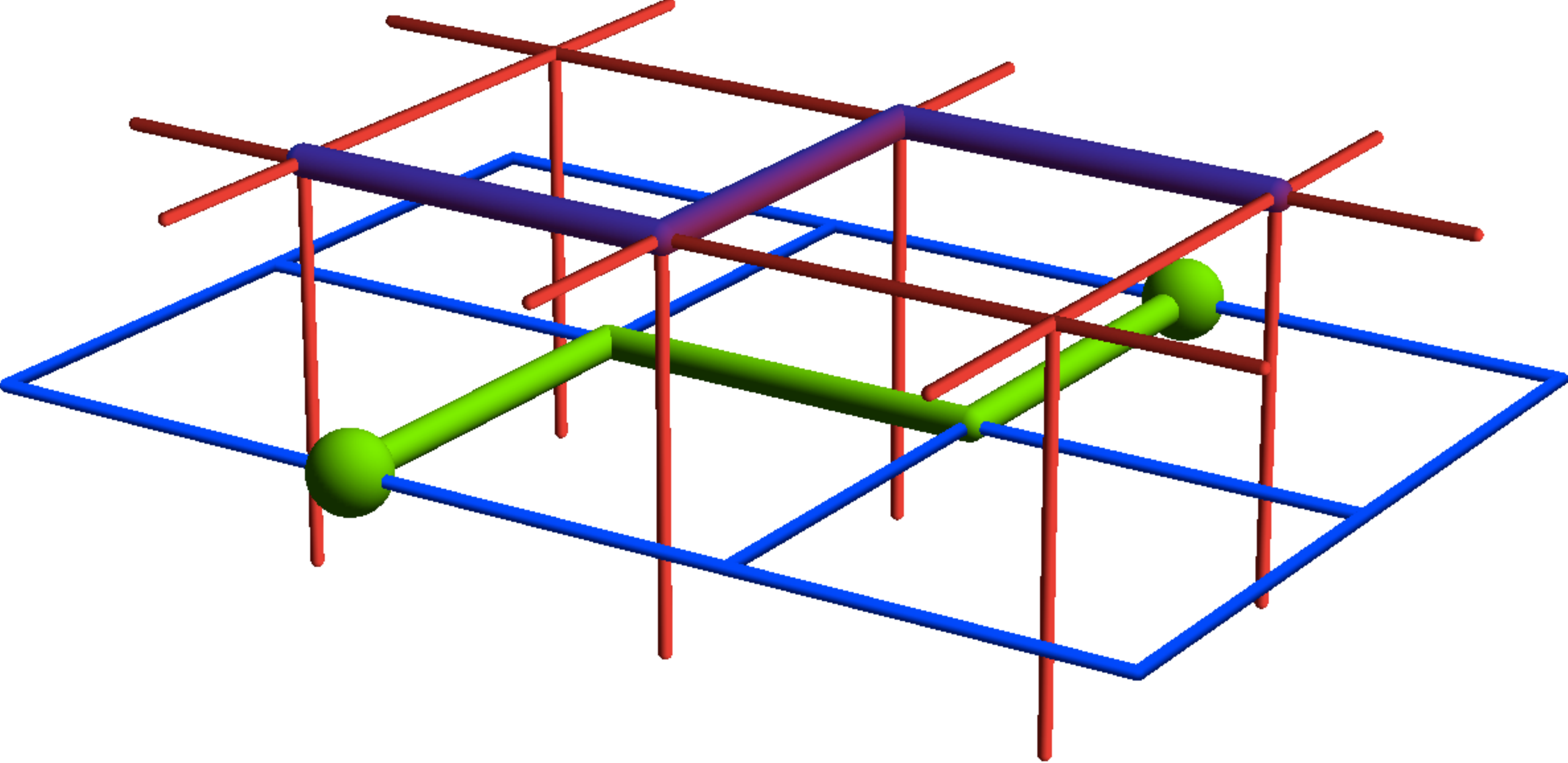}
\caption{String operators creating surface excitations. Left:  acting with a chain of $\sigma^x$ operators on the links of the upper layer ($A$ lattice surface) gives a pair of $e$ excitations (i.e. endpoints of bulk $A$ strings). Right: a pair of $m$ excitations (i.e. endpoints of $B$ strings) are created by a chain of $\tau^x$ operators (thick green strand) on the lower layer ($B$ surface), together with $\sigma^z$ operators on the corresponding links in the upper layer (thick purple links).}
\label{surfaceexcitations}
\end{figure}

Despite the lack of deconfined endpoints in the bulk,  $A$ and $B$ strings that terminate on a boundary can give deconfined $e$ and $m$ particles in a surface $\mathbb{Z}_2$ topologically ordered state. To see this, we terminate the system as in Fig.~\ref{surfaceexcitations}, including in the Hamiltonian the natural plaquette and vertex terms at the surface. The surface string operators that create pairs of  $e$ or pairs of  $m$ excitations can then be written explicitly (see Fig.~\ref{surfaceexcitations}). They satisfy the same algebra as the string operators in the toric code \cite{kitaev toric code}, confirming that $e$ and $m$ are mutual semions as expected from the heuristic argument of Fig.~\ref{surface braiding}.

We can adiabatically transform $\ket{\Psi}$ to a product state so long as we allow the intermediate states to violate the closed--loop constraints on at least one sublattice. The  membrane picture gives an obvious way to do this, by giving the membranes in $\mathcal{M}$ a surface tension. If `Area' denotes the number of occupied plaquettes in $\mathcal{M}$,  the interpolating state is
\be\label{interpolating state}
\ket{\Psi}_\gamma =
\sum_\mathcal{\CC_A} 
\sum_{\substack{
\mathcal{M} \\
\partial \mathcal{M} = \CC_A }}
e^{ -\gamma \times \text{Area} }
 \ket{\CC_A}
\ket{\mathcal{M}}.
\ee
When $\gamma=0$ this is the initial state, and when $\gamma\rightarrow\infty$ only the term with zero area survives. This is the state with no loops and no membranes, i.e. the product state $\ket{\sigma^z = 1}\ket{\tau^x = 1}$. To get a gapped parent Hamiltonian for $\ket{\Psi}_\gamma$, we modify the plaquette flip term $\FF_A$ in  $\HH_\text{linking}$ to $\FF_{Ap}=(\cosh \gamma)^{-1} \Big[ \tau^z_p \prod_{l\in p} \sigma_l^x + (\sinh \gamma) \, \tau^x_p \Big]$. This preserves the simple algebraic properties  of the plaquette terms.

\subsection{Self-duality of $\ket{\Psi}$ and protection by constraints}
\label{self-duality}

When the interpolating state above is rewritten in the original $(\sigma^z, \tau^z)$ basis, it includes configurations with open strands, as well as closed loops, on the $B$ lattice. What if we impose the  constraint that both lattices have only closed loops? In this case it is impossible to go from $\ket{\Psi}$ to a trivial state  without a phase transition. (We will take the reference trivial state to be that with no loops, $\ket{\text{trivial}} = \ket{\sigma^z = 1}\ket{\tau^z = 1}$.)

This follows from a simple duality transformation which exchanges the electric flux of each species with the magnetic flux of the other species. The duality maps $\ket{\Psi}$ to itself, but exchanges the trivial state with a topologically ordered one. Thus there is no adiabatic path from $\ket{\Psi}$ to the trivial state. If there were, duality would yield an adiabatic path from $\ket{\Psi}$ to the topologically ordered state, and this is impossible since $\ket{\Psi}$ is not topologically ordered.

\begin{figure}[t] 
\centering
\includegraphics[width=2.8in]{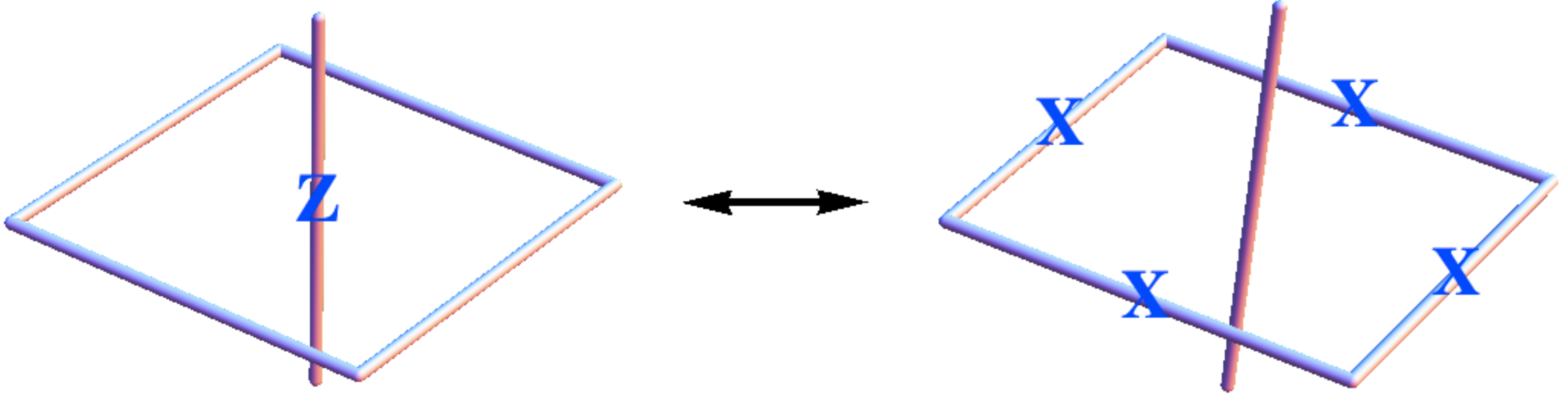}
\caption{Under the mapping (\ref{operatormapping}), a $\sigma^z$ (or $\tau^z$) operator on a link is exchanged with a product of $\tau^x$ (resp. $\sigma^x$) operators on the surrounding links of the other lattice. (Links of one lattice can equally be thought of as plaquettes of the other.)}
\label{dualitypic}
\end{figure}

The duality transformation makes sense for states obeying the closed loop constraint. (To be precise, we must also impose the global constraint that the loop configurations have even winding in each direction.) As shown in Fig.~\ref{dualitypic}, its action is:
\ba\label{operatormapping}
\sigma_l^z & \longleftrightarrow \prod_{p\in l} \tau^x_p, &
\tau^z_p & \longleftrightarrow \prod_{l\in p} \sigma^x_l 
\end{align}
Here $p\in l$ denotes the four plaquettes $p$ surrounding link $l$. We have labelled the $\sigma$s by $l$ for link and the $\tau$s by $p$ for plaquette, but the duality acts on the two sets of degrees of freedom symmetrically. It preserves the locality of any Hamiltonian acting in the constrained Hilbert space.

For completeness, we write the action of the duality on states explicitly. Return to the picture of loops + membranes on the $A$ lattice, i.e. the $(\sigma^z, \tau^x)$ basis. One may check that any state satisfying the constraints  can be written as a sum over two loop configurations on the \emph{same} lattice,
\be\label{general constrained ket}
\ket{f} = \sum_{\CC_A, \CC_A'} f(\CC_A, \CC_A') \ket{\CC_A}_{\sigma} \widetilde{\ket{\CC_A'}}_{\tau},
\ee
where $\widetilde{\ket{\CC_A'}}_{\tau}$ is defined as the uniform superposition of all membrane configurations $\ket{\mathcal{M}}_{\tau}$ with boundary $\partial \mathcal{M} =\CC_A'$. We have added subscripts to the kets as a reminder of the degrees of freedom involved. ($\CC_A$ is the $\sigma$--electric flux configuration, and $\CC_A'$ the $\tau$--magnetic flux configuration; the fact that the wavefunction depends on $\mathcal{M}$ only through $\partial \mathcal{M}$ is simply a statement of gauge invariance.) The duality then simply exchanges the two kinds of loops,
\be\label{action of duality on wavefunction}
f(\CC_A, \CC_A')\longleftrightarrow f(\CC_A', \CC_A).
\ee

The flip operators $\FF_A$ and $\FF_B$ (Eq.~\ref{flip operators}) are clearly invariant under the duality in Eq.~\ref{operatormapping} and therefore so is $\HH_\text{linking}$. (We can also see that $\ket{\Psi}$ is invariant from Eq.~\ref{action of duality on wavefunction} and Eq.~\ref{surface representation of pure loop state}.) On the other hand, the trivial Hamiltonian 
\be
\HH_\text{trivial} = - \lf J \sum_{l\in A} \sigma^z_l + J \sum_{l\in B} \tau^z_l \ri
\ee
is exchanged with
\be
\HH_\text{deconfined} = - \lf J  \sum_{p\in A} \prod_{l\in p} \sigma^x_l - J  \sum_{p\in B} \prod_{l\in p} \tau^x_l \ri,
\ee
which describes a pair of deconfined $\mathbb{Z}_2$ gauge theories. This establishes the claim at the beginning of this subsection: while the linking state is invariant, the trivial state is exchanged with a topologically ordered state. It follows that the linking state is in a distinct phase from the trivial state if we do not allow open endpoints in the Hilbert space.  (We know from Eq.~\ref{interpolating state} that they are in the same phase if we \emph{do} allow endpoints.)

\subsection{Heuristic relation between symmetry protection of $eTmT$ and closed--loop constraint}

The proposed wave function for the $eTmT$ phase has the two loop species `stuffed' with Haldane/AKLT chains. The linking sign factor ensures that the ground state is not topologically ordered as required for a topological paramagnet. In particular the open end-points of the loops --- which now harbor a Kramers doublet --- are confined. However as described in Sec. \ref{fluctuating AKLT chains} the surface implements time-reversal  `anamolously'  exactly characteristic of the $eTmT$ state. 

 We now briefly consider whether the results in the previous subsection for the `pure loop' state yield a heuristic `bulk' understanding of why the $eTmT$ state is protected by time reversal.   So let us imagine perturbing the schematic $eTmT$ wavefunction of Sec.~\ref{fluctuating AKLT chains}, and ask why we cannot reach a trivial state without a phase transition.

We make use of the heuristic analogy between the AKLT loops of the spin-1 system and the `pure loops' of the coupled gauge theory.\footnote{Recall that the singlet basis allows us to represent any spin-zero state of the  spin-1 system in terms of loops of {spin-1/2} singlet bonds; these may form minimal length `loops' which backtrack on a single link, i.e. spin-1 {singlet bonds}, or longer AKLT loops.} The  result for the pure loop state then indicates that if we only have closed AKLT loops on each sublattice, we cannot get to a trivial state without a phase transition. So, we must consider proliferating open strands on at least one sublattice. But in the spin-1 system, unlike the pure-loop system,  open strands introduce bulk spin-1/2 Kramers doublet degrees of freedom. (Binding these emergent spin--1/2s into singlets with others on the {same} sublattice merely heals the AKLT chains, taking us back to the original situation with separate closed loops on each sublattice.)
When time reversal is broken, these spin--1/2s are innocuous --- for example we can gap them out using a magnetic field. But it is natural to expect that when time reversal is preserved they prevent us reaching a trivial state without closing the gap.

However, the above argument is incomplete as it does not rule out the possibility of getting to a trivial state by proliferating nearby pairs of open strands on \emph{opposite} sublattices. Such a pair gives two spin--1/2s which can be bound into a singlet to avoid a gapless degree of freedom. In the gauge theory, such pairs correspond to bound pairs of electric charges, one from each $\mathbb{Z}_2$ gauge field.  The stability of the $eTmT$ state suggests that the pure loop state remains protected even when such double charges are allowed.  We note that at the surface these double charges correspond to the bound state of the $e$ and $m$ particle (in the surface topological order). This is a Kramers singlet spin-$0$ fermion (conventionally denoted $\epsilon$).  The surface Fermi statistics suggests a potential obstruction to `trivializing' the bulk by proliferating the double charges. We leave an explicit demonstration of this for the future.

\section{Parton constructions}
Though the description of the $eTmT$ topological paramagnet in terms of a loop gas wave function is physically appealing it is desirable to have alternate descriptions which enhance our understanding and which may help with evaluating the energetic stability of this phase in microscopic models. 
To that end, in this section we propose explicit parton constructions for some topological paramagnets in spin-$1$ systems.

Historically the parton approach has provided variational wave functions and effective field theories both for  spin liquids \cite{Wenbook} and non-fractionalized symmetry-breaking states \cite{auerbach}. The parton construction inevitably introduces a gauge symmetry. It describes a fractionalized spin liquid phase whenever it yields an emergent deconfined gauge field. To obtain a non-fractionalized phase such as conventional antiferromagnet or valence bond solid paramagnet, the gauge field should either be Higgsed or confined. 

Recently the parton construction has been used to construct SPT states in two \cite{2dparton,cenkeprtn2d} and three \cite{3dpartonu1,3dpartonz2} dimensions. The general idea is to construct a gauge theory (with matter fields) that is confined, but with certain non-trivial features surviving in the confined state that make it an SPT state. However, the currently known constructions in three dimensions use either $\mathbb{Z}_2$ or $\mathrm{U}(1)$ gauge theories, which do not confine automatically: strong gauge coupling is needed to reach the confined phase. Furthermore, the constructions using $\mathrm{U}(1)$ gauge theories \cite{3dpartonu1} require highly nontrivial dynamics of the gauge fields to condense composite dyon-like objects. 

In three dimensions, a continuous non-abelian gauge symmetry is needed to guarantee confinement.  
We propose two parton constructions in three dimensions with $\mathrm{SU}(2)$ gauge symmetry, which confine even if the bare gauge coupling is small, giving rise to topological paramagnets. A similar construction was used previously \cite{cenkeprtn2d} in 2D to describe an SPT phase of a spin-$1$ magnet protected by spin $\mathrm{SU}(2)$ symmetry and time reversal. 
We also propose a  construction with $\mathrm{U}(1)$ gauge symmetry, which confines at sufficiently strong coupling. Crucially, this $\mathrm{U}(1)$ construction differs from previous ones in that we only condense simple monopoles to confine the gauge theory, which can be achieved at strong coupling without exotic form of gauge field dynamics.

The spin-$1$ operators are re-written using the two-orbital fermionic parton representation proposed in Ref.~\onlinecite{cenkeetal},
\be
\label{slave}
\vec{S}=\frac{1}{2}\sum_{a=1,2}f^{\dagger}_{a\alpha}\vec{\sigma}^{\phantom{\dag}}_{\alpha\beta}f^{\phantom{\dag}}_{a\beta}.
\ee
where $a=1,2$ is the orbital index. As will be discussed below, the two-orbital structure is natural for topological bands corresponding to topological paramagnets. This gives another reason for favoring spin-$1$ systems.

The physical spin states are represented in the parton description as
\bea \notag
&|S_z=0\rangle=\f{1}{\sqrt 2} \lf f^{\dagger}_{1\down}f^{\dagger}_{2\up}+f^{\dagger}_{1\up}f^{\dagger}_{2\down} \ri |\text{vac}\rangle, \nn
&|S_z=+1\rangle=f^{\dagger}_{1\up}f^{\dagger}_{2\up}|\text{vac}\rangle, \quad\quad
|S_z=-1\rangle=f^{\dagger}_{1\down}f^{\dagger}_{2\down}|\text{vac}\rangle.
\eea
where $|\text{vac}\rangle$ is the state with no fermions. States in the physical spin Hilbert space thus have two fermions at each site, $\sum_{a\alpha} f^{\dagger}_{a\alpha}f_{a\alpha}=2$, and the two fermions form a singlet in orbital space:   denoting the Pauli matrices in orbital space by $ \tau^{x,y,z}$, this is $\sum_{\alpha}f^{\dagger}_{a\alpha}\vec \tau_{ab}f_{b\alpha}=0$.

The representation in Eq.~\ref{slave} actually has an $\SP(4)$ gauge redundancy \cite{cenkeetal}  which becomes apparent when we represent the fermions using  Majoranas,  $f=\f{1}{2} \lf \eta_1- i\eta_2 \ri$. Here $\eta_{1,2}$ are Hermitian operators satisfying $\{\eta_{sI},\eta_{s'J}\}=2\delta_{ss'}\delta_{IJ}$, where $s,s'=1,2$ are the new indices associated with the Majoranas and $I,J$ represent all other indices (site, spin, orbital). The Majorana representation of the spin is
\ba
\label{slave2}
\vec{S}& =\frac{1}{8}\eta^T \vec \Sigma \, \eta, & 
\vec \Sigma &= \lp\rho^y\sigma^x,\sigma^y,\rho^y\sigma^z\rp,
\end{align}
where $\rho^{x,y,z}$ are Pauli matrices acting on the Majorana index. The generators of the gauge symmetry are ten anti-symmetric imaginary matrices that commute with the physical spin operators:
\be\label{list of gammas}
\Gamma=\{\rho^y,\rho^y\tau^{x,z},\rho^{x,z}\sigma^y,\rho^{x,z}\sigma^y\tau^{x,z},\tau^y\},
\ee
where $\tau_i$ are Pauli matrices acting on the orbital index. The spin in Eq.~\ref{slave2} is invariant under the $\SP(4)$ gauge transformation $\eta\to e^{ia_i\Gamma_i}\eta$.
 
The effective field theory associated with the parton construction is a gauge theory. The gauge symmetry is determined by the mean field band structure of the partons, and is in general a subgroup of the full $\SP(4)$ group due to some generators being Higgsed. The gauge structure allows symmetry to act projectively on the $\eta$ fermion \cite{Wenbook}. In particular, time-reversal could be either Kramers ($\mathcal{T}^2=-1$) or non-Kramers ($\T^2=1$). 

In 3D, band structures of Kramers fermions with $\mathcal{T}$ symmetry are classified by an integer index \cite{tenfold} $\nu$ which counts the number of Majorana cones on the surface. It was realized \cite{fidkowski3d,3dfSPT2,maxvortex} that in the presence of interactions the state with $\nu=16$ is trivial, while that with $\nu=8$ is equivalent to a topological paramagnet. More specifically, for $\nu=8$ the surface state with four Dirac cones (eight Majorana cones) can be gapped without breaking any symmetry via strong interactions, and the resulting gapped surface state must have intrinsic topological order. The simplest such topological order is a $\mathbb{Z}_2$ gauge theory in which the $e$ and $m$ particles are bosons, but transform under time-reversal as Kramers doublets ($\T^2=-1$). Therefore we can put the slave fermions into a band with $\nu=8$, and let the gauge fields confine the fermions (either automatically through an $\mathrm{SU}(2)$ gauge field or at strong coupling through a $\mathrm{U}(1)$ gauge 
field). 
Crucially, the topological quasi-particles ($e$ and $m$) on the surface do not carry the gauge charge, and they survive on the surface as deconfined objects.
The resulting phases 
are therefore confined paramagnets with nontrivial surface states protected by time-reversal symmetry.

Non-Kramers fermions, by contrast, cannot host non-trivial band structure with time-reversal symmetry alone. However, if spin--$S_z$ conservation is present, the band structures can again be assigned an integer topological invariant $\nu'$ \cite{tenfold} which is the number of Dirac cones on the surface (or half the number of Majorana cones). It is known \cite{fidkowski3d,3dfSPT2,maxvortex} that with interactions the state with $\nu'=8$ is trivial, while that with $\nu'=4$ is equivalent to a topological paramagnet. We can then put the slave-fermions into a band with $\nu'=4$ and let the gauge fields confine the fermions, which produces a topological paramagnet with time-reversal and spin-$S_z$ conservation.

In both cases we need to put the slave fermions into band structures with four Dirac cones on the surface. Band structures with two Dirac cones ($\nu = 4$) have been studied on the cubic \cite{hrv} and diamond \cite{srl} lattices. Therefore we can obtain the desired structure  simply by putting the partons into two copies of the $\nu=4$ band. This can be easily done by taking advantage of the two orbitals in Eq.~\ref{slave},  making the topological paramagnets very natural in spin-$1$ systems.

In the next section we outline a similar construction for the one--dimensional Haldane chain, by confining slave fermions which form four copies of the Kitaev chain. This illustrates the essential idea of our constructions in a simpler and more familiar context.

\subsection{Parton construction for Haldane/AKLT chain}

The Haldane phase is an SPT phase with gapless boundary degrees of freedom that are protected by time reversal. As a warm--up exercise, we outline how this phase can be constructed from a topological superconductor of slave fermions. This illustrates some features we will meet again in 3D. A different parton construction for the Haldane phase was considered in Ref.~\onlinecite{1dparton}.

The fermions are taken to be non-Kramers ($\T^2 = 1$). In 1D, superconducting band structures for 
\emph{free} non-Kramers fermions are labelled by a $\mathbb{Z}$--valued index \cite{tenfold}, $\nu$, which is the number of protected Majorana zero modes at the boundary. The state with a given $\nu$ can be viewed as $\nu$ copies of Kitaev's p--wave superconducting chain \cite{Kitaev chain}. Interactions reduce this classification to $\mathbb{Z}_8$, i.e. the $\nu = 8$ state becomes trivial \cite{1dz8}. Further, the state with $\nu = 4$ is topologically equivalent  to the Haldane chain, modulo the presence of gapped fermions in a trivial band. 

Here we therefore put the slave fermions into four copies of the Kitaev bandstructure, in an $\mathrm{SU}(2)$--symmetric manner. Gauge fluctuations (or Gutzwiller projection) will then remove the unwanted degrees of freedom, leaving a topological paramagnet in the Haldane phase.

Starting with an antiferromagnetic spin--1 chain,
\be\label{H 1D}
\HH = J \sum_i \vec S_i . \vec S_{i+1}  + \ldots,
\ee
we represent the spins with slave fermions as in Eq.~\ref{slave} or equivalently Eq.~\ref{slave2}.  The valence bond picture of the AKLT state suggests using a mean-field Hamiltonian for the partons with hopping $t$ and spin--singlet, orbital--singlet pairing $\Delta$,
\ba\notag
H_\mathrm{MF} & = 
- \sum_i \left[ 
t \left[ f_i^\dag f^{\phantom{\dag}}_{i+1} + \text{h.c.} \right] 
+ \Delta \left[ f_i^\dag \sigma^y \tau^y f_{i+1}^{\dag T} +\text{h.c.} \right]
\right].
\end{align}
In terms of the Majoranas, this is
\ba\label{H 1D MF}
H_\mathrm{MF} & = -\f{1}{2} \sum_i \eta_i^T M \eta^{\phantom{T}}_{i+1}, & M& = t \rho^y +i \Delta \rho^x \sigma^y \tau^y.
\end{align}
We first consider this as a free fermion problem,  then include the gauge fluctuations.

For simplicity take $\Delta=t$, which makes the terms in $H_\text{MF}$ for different links commute. The Hamiltonian is simply four copies of the Kitaev chain, as can be seen immediately by going to a basis where $\sigma^y \tau^y$ is diagonal.  To be more explicit, it is useful to define the matrix 
\be\label{X matrix}
X=\rho^z \sigma^y \tau^y.
\ee
Firstly, we use this to define the action of time reversal $\T$ on the fermions:
\be
\T: \qquad \eta \longrightarrow X \eta.
\ee
This definition ensures that the spin changes sign under $\T$ and that $H_\mathrm{MF}$ is invariant. The fermions are non-Kramers ($\T^2 =1$ on $\eta$).

Secondly let us define matrices that project onto a given value of $X$, and corresponding fermion modes:
\ba
\mathcal{P}_\pm & = \f{1}{2}(1\pm X), &
\eta^{(\pm)} = \mathcal{P}_\pm \eta.
\end{align}
In an appropriate basis, $\eta^{(+)}$ has four nonzero components. Next, note that
\be
M = \mathcal{P}_- M \mathcal{P}_+,
\ee
since $M  =  t \rho^y (1+ \rho^z \sigma^y \tau^y) = (2 t \rho^y) \mathcal{P_+}$, etc. So we may rewrite $H_\mathrm{MF}$ as
\ba
H_\mathrm{MF} & = -\f{1}{2} \sum_i \eta_i^{(-)T} M \eta^{(+)}_{i+1}.
\end{align}
Taking open boundary conditions, and denoting the leftmost site of the chain by $L$, we see that the four modes in $\eta^{(+)}_L$ do not appear in the Hamiltonian. 

These four Majoranas correspond to two complex fermion modes that can be occupied or unoccupied, i.e to a degenerate four-dimensional boundary Hilbert space.  At the level of free fermions, this degeneracy is protected by time reversal symmetry $\T$, under which $\eta^{(+)}_L$ is invariant (since by definition  $X\eta^{(+)} = \eta^{(+)}$).\footnote{Any quadratic term $i \eta^{(+)T} A \eta^{(+)}$ (where $A$ is real antisymmetric) is forbidden as it is odd under $\T$. However in the presence of interactions a four fermion term $\gamma_1\gamma_2\gamma_3\gamma_4$ --- where $\gamma_i$ are the components of $\eta^{(+)}$ in some basis --- is allowed by time reversal, and lifts the boundary degeneracy to a single doublet as in the Haldane chain \cite{tangwen}.}

Once we go beyond mean field theory, the fermions are coupled to confining gauge fluctuations. We will see below that two of the four boundary states are not gauge invariant --- i.e. they can be thought of as having an unscreened gauge charge sitting at the end of the chain.  Confinement removes these states from the low energy Hilbert space, leaving a single boundary spin--1/2 whose gaplessness is protected by time reversal.

$H_\mathrm{MF}$ treats spin and orbital degrees of freedom symmetrically, and preserves $\mathrm{SU}(2)_\text{spin}\times\mathrm{SU}(2)_\text{orbital}$ symmetry. 
The four boundary states can be labeled by the occupation numbers of two complex fermions $c_{1,2}$. Since the partons transform as doublets under each $\mathrm{SU}(2)$, the fermions $c_{1,2}$ should also form doublets under each $\mathrm{SU}(2)$. In an appropriate basis the transformations are
\ba
&\mathrm{SU}(2)_\text{spin}: &   (c_1,c_2)^T&\,\longrightarrow\,\mathcal{U}_s(c_1,c_2)^T, \nonumber \\
&\mathrm{SU}(2)_\text{orbital}:&  (c_1,c_2^{\dagger})^T&\,\longrightarrow\,\mathcal{U}_o(c_1,c_2^{\dagger})^T.
\end{align}
where $\mathcal{U}_{s,o}$ are $\mathrm{SU}(2)$ matrices. It follows that states which are singlets under $\mathrm{SU}(2)_\text{spin}$ are doublets under $\mathrm{SU}(2)_\text{orbital}$ and vice versa.  We denote the spin doublet  $\ket{\uparrow}, \ket{\downarrow}$ and the orbital doublet $\ket{1}$, $\ket{2}$. The spin operator for the boundary spin-1 can be split into contributions from the dangling boundary modes $\eta^{(+)}_L$ and from $\eta^{(-)}_L$: $\vec S_L^{\phantom{+}} = \vec S_L^{(+)} + \vec S_L^{(-)}$, with
\ba
\vec S^{(\pm)} &= \f{1}{8} \eta^{(\pm)T} \vec \Sigma \, \eta^{(\pm)}, & 
\vec \Sigma & = (\rho^y \sigma^x, \sigma^y, \rho^y\sigma^z).
\end{align}
We can make a similar splitting for the orbital spin $\vec T$, which is related to $\vec S$ by swapping the $\sigma$s for $\tau$s. We denote the matrices appearing in $\vec T$ by $\vec\Omega$:
\ba\label{orbital spin}
\vec T^{(\pm)} &= \f{1}{8} \eta^{(\pm)T} \vec \Omega \, \eta^{(\pm)}, & 
\vec \Omega & = (\rho^y \tau^x, \tau^y, \rho^y\tau^z).
\end{align}  
The pairs $(\ket{\uparrow}, \ket{\downarrow})$ and $(\ket{1},\ket{2})$ are both Kramers doublets, since the spin and orbital operators for the boundary modes, $\vec S_L^{(+)}$ and $\vec T_L^{(+)}$,  change sign under $\T$. This can also be checked explicitly by considering the transformation of the boundary states (labeled by fermion occupation numbers) under $\T$, with the fermions transforming as $\T: c_{1,2}\to c^{\dagger}_{1,2}$.

Now we consider the effect of gauge fluctuations or Gutzwiller projection. We have listed  the generators for the $\mathrm{Sp}(4)$ gauge group in Eq.~\ref{list of gammas}. However, some gauge generators are Higgsed in the above mean field state. In general, to determine the unbroken gauge group we must examine Wilson loops of the form $W=\hat{u}_{i_1i_2}\hat{u}_{i_2i_3}..\hat{u}_{i_ni_1}$, where $H_{\text{MF}}=\sum_{ij}\eta^T_i\hat{u}_{ij}\eta_j$ \cite{Wenbook}. The unbroken gauge generators are those that commute with the Wilson loops. Here, the only nontrivial Wilson loop is the matrix $X$ defined in Eq.~\ref{X matrix}. This leaves a subset of six unbroken generators, which may be written in terms of the matrices $\Omega$  appearing in the orbital spin (Eq.~\ref{orbital spin}):
\be\label{gamma 1D}
\Gamma_\text{1D} = \{ \vec \Omega, X \vec \Omega \}.
\ee
Taking linear combinations, we can use instead\footnote{To be more precise, the two types of generators in Eq.~\ref{gamma 1D} correspond to elements of the invariant gauge group (IGG) \cite{Wenbook} at different momenta, $k=0$ and $k=\pi$, so when we take linear combinations  the two types of generators in Eq.~\ref{gamma 1D 2} alternate on even and odd sites. This is not crucial here. Another subtlety is that the IGG is enlarged at the special point $\Delta = t$.}
\be\label{gamma 1D 2}
\Gamma_\text{1D} = \{ \mathcal{P}_+ \vec \Omega \mathcal{P}_+, 
\mathcal{P}_- \vec \Omega \mathcal{P}_-\}.
\ee
We denote the unbroken gauge group $\mathrm{SU}(2)^{(+)}_\text{orbital}\times \mathrm{SU}(2)^{(-)}_\text{orbital}$.

To make the Hamiltonian in Eq.~\eqref{H 1D MF} a reasonable ansatz, we must check that the $\SP(4)$ gauge charges are all zero on average: $\langle\Gamma_i\rangle=0$ for all $i$. Fortunately the unbroken gauge symmetry $\Gamma_\text{1D}$ guarantees this. 

The boundary modes involve only $\eta^{(+)}$, so are invariant under $ \mathrm{SU}(2)^{(-)}_\text{orbital}$. However, $\ket{1}$ and $\ket{2}$ are not invariant under $SU(2)^{(+)}_\text{orbital}$. Therefore after confinement only the doublet $\ket{\uparrow}$, $\ket{\downarrow}$ survives, with corresponding spin $\vec S^{(+)}_L$. This is the boundary spin-1/2 of the Haldane phase.

In this 1D example we can confirm explicitly that Gutzwiller--projecting the mean-field wavefunction gives the desired SPT phase. In fact  the Gutzwiller--projected state for $\Delta = t$, denoted $\ket{\Psi_\text{spin}}$, is precisely the AKLT state. To see this we adopt a trick from Ref.~\onlinecite{1dparton}. Using the fact that the terms in $H_\text{MF}$ commute, we can check that $\ket{\Psi_\text{spin}}$ has zero amplitude for a pair of adjacent sites to be in a spin--two state. $\ket{\Psi_\text{spin}}$ is therefore the ground state of the AKLT Hamiltonian, since this can be written as a sum of projectors onto the spin-two subspace for each link.\footnote{This correspondence with the AKLT state is less obvious if we simply Gutzwiller--project the BCS ground state of $H_\text{MF}$. The ground state of the Kitaev chain involves a long-range Cooper pair wavefunction, $C(r) = (L-2r)/L$ \cite{Greiter Schnells Thomale}, so in the present case $\ket{\Psi_\text{spin}}$ is obtained by acting on the vacuum with an exponented sum of long-range singlet creation operators, $\exp (\sum_i \sum_{r>0} C(r) f_i^\dag \sigma^y \tau^y f_{i+r}^{\dag T})$, and projecting. The AKLT state may of course be written using only short--range singlet creation operators, $\ket{\text{AKLT}} \propto \mathcal{P} \prod_i ( f_i^\dag \sigma^y \tau^y f_{i+1}^{\dag T} ) \ket{\mathrm{vac}}$. By the previous argument, the two states must be equivalent.}

It is interesting to consider inversion symmetry here. In the free fermion problem, $\nu\rightarrow -\nu$ under inversion, so that a nonzero value of $\nu$ can only be realised with a Hamiltonian which breaks inversion symmetry. With interactions, $\nu \simeq \nu + 8$, suggesting that $\nu = 4$ can be realised in inversion--symmetric interacting system \cite{hughes}. The present example is a nice realisation of this. The mean field Hamiltonian $H_\mathrm{MF}$ appears to break inversion symmetry. However, the symmetry can be restored by combining it with a gauge rotation. So the projected wavefunction is actually inversion symmetric.

We now move on to 3D states.

\subsection{Cubic lattice}
\label{cube}

Making use of the cubic band structure studied in Ref.~\onlinecite{hrv}, we construct an $\mathrm{SU}(2)$ gauge theory which confines to a topological paramagnet.  We choose the mean field Hamiltonian
\bea
\label{cubband}
H_{\text{MF}}=&&\sum_{\langle ij\rangle}t_{ij}\eta_i^{T}\rho^y\eta_j+\sum_{\langle\langle ij\rangle\rangle}i\chi'_{ij}\eta_i^T\rho^x\sigma^y\tau^y\eta_j \nn
&&+\sum_{\langle\langle\langle ij \rangle\rangle\rangle}\chi_{ij}\eta^T_i\rho^x\sigma^y\eta_j,
\eea
where the nearest-neighbor hopping $t_{ij}$ gives a $\pi$-flux on every square plaquette, the body-diagonal pairing $\chi_{ij}$ follows the pattern studied in Ref.~\onlinecite{hrv}, and the next-nearest-neighbor pairing $\chi'_{ij}$ is a small perturbation introduced to reduce the gauge group to $\mathrm{SU}(2)$ and is not responsible for the gap or the band topology.

To determine the unbroken gauge group, we examine the Wilson loops  as above. The fundamental nontrivial ones are proportional to $\rho^z\sigma^y$ and $\rho^x\sigma^y\tau^y$. The unbroken gauge group is generated by those of the $\mathrm{Sp}(4)$ generators that commute with the Wilson loops. It is then straightforward to see that the unbroken gauge group is an $\mathrm{SU}(2)$ generated by 
\be\label{cubic gauge generators}
\Gamma_\text{cubic}=\{ \rho^z\sigma^y\tau^x,\tau^y,\rho^z\sigma^y\tau^z \}.
\ee

One can choose to implement time-reversal $\T$ as $\T: \eta\to i\rho^z\sigma^y\eta$, and it is straightforward to see that $\T: H_{\text{MF}}\to H_{\text{MF}},  \vec{S}\to-\vec{S}$, and $\vec{\Gamma}_\text{cubic} \to-\vec{\Gamma}_\text{cubic}$.  The band structure in Eq.~\eqref{cubband}  preserves time-reversal symmetry, and the $\mathrm{SU}(2)$ gauge rotation commutes with $\T$. Notice also that $\T^2=-1$ on the $\eta$ fermions.

We must check that the $\SP(4)$ gauge charges are all zero on average, $\langle\Gamma_i\rangle=0$. The unbroken gauge symmetry $\Gamma_\text{cubic}$ guarantees that $\langle\Gamma_i\rangle=0$ for all $i$ except for $\Gamma_5=\rho^z\sigma^y$. Furthermore, time-reversal invariance $\T$ guarantees that $\langle\Gamma_5\rangle=0$. Hence the condition is indeed satisfied for any $i$.

To determine the band topology, it is sufficient to consider the Hamiltonian $H'_{\text{MF}}$ with only the nearest-neighbor and body-diagonal terms in Eq.~\eqref{cubband}. In $H'_{\text{MF}}$, fermions with different orbital indices are decoupled and form two identical bands. Each band is the same as that studied in Ref.~\onlinecite{hrv}, with $\nu=4$ (two Dirac cones on the surface). So the band has $\nu=8$ in total (four Dirac cones). So Eq.~\eqref{cubband} indeed gives rise to a topological paramagnet.

In order to understand the role played by spin-rotation symmetry, we examine the surface state in more detail. We start from the surface Dirac theory with~$\mathrm{SU}(2)_\text{gauge}\times~\mathrm{SU}(2)_\text{spin}\times\mathcal{T}$ symmetry, with four Dirac cones in total:

\be\label{surface dirac hamiltonian}
H=\psi^{\dagger}(p_x\mu_x+p_y\mu_z)\otimes\tau_0\otimes\sigma_0\psi,
\ee
with time-reversal
\be
\mathcal{T}:\psi\to i\mu_y\otimes\tau_0\otimes\sigma_0\psi^{\dagger},
\ee
gauge $\mathrm{SU}(2)$
\be
\mathcal{U}_g: \psi\to\mu_0\otimes \mathcal{U}_g\otimes\sigma_0\psi,
\ee
and spin $\mathrm{SU}(2)$
\be
\mathcal{U}_s: \psi\to\mu_0\otimes\tau_0\otimes \mathcal{U}_s\psi.
\ee
We have denoted the $\mathrm{SU}(2)_\text{gauge}$ Pauli matrices by $\vec\tau$, but they  should not be confused with the Pauli matrices for the orbital spin. 

Next we will consider driving this surface theory into a $\mathbb{Z}_2$ topologically ordered state by first introducing an order parameter $\Delta$ which gaps out the Dirac fermions, but breaks time reversal symmetry, and then restoring time--reversal symmetry by proliferating double vortices in $\Delta$. The single vortex remains gapped, and gives rise to anyonic surface excitations with nontrivial time reversal properties.

To analyse the symmetry properties it is useful to consider the auxiliary $\mathrm{U}(1)_a$ transformation
\be
U_a(\theta): \psi\to e^{i\theta}\psi
\ee
(which is an emergent symmetry of Eq.~\ref{surface dirac hamiltonian}, but not a microscopic symmetry). The gap term of interest is
\be
H_{\Delta}=i\Delta\,\psi\mu_y\otimes\tau_y\otimes\sigma_y\psi+\mathrm{h.c.}
\ee
This is invariant under the $\mathrm{SU}(2)_\text{gauge} \times \mathrm{SU}(2)_\text{spin}$ symmetry. It is not invariant under time reversal $\T$ or under $\mathrm{U}(1)_a$ separately, but it   is invariant under the modified time-reversal transformation $\tilde{\T}\equiv U_a(\pi/2)\T$. Notice that $\tilde{\T}^2=1$ on the parton fermions $\psi$,   in contrast to the original $\T$ under which they are Kramers.  

As shown in Refs.~\onlinecite{wpssc14,3dfSPT2,maxvortex}, the fundamental vortex in $\Delta$ transforms projectively under $\tilde{\T}$, i.e.  $\tilde{\T}^2=-1$.  We now examine the $\mathrm{SU}(2)_\text{gauge} \times \mathrm{SU}(2)_\text{spin}$ spins carried by the vortex. A key point is that there are four Majorana zero modes trapped in the vortex core. One can label the internal Hilbert space with two complex fermions $c_{1,2}$. Since both $\mathrm{SU}(2)$ groups are preserved in the intermediate gapped phase and the partons transform as doublets under both $\mathrm{SU}(2)$, the two complex fermions $c_{1,2}$ should also be doublets under both $\mathrm{SU}(2)$. In an appropriate basis the transformations are
\bea
\mathcal{U}_g:&& (c_1,c_2)^T\to\mathcal{U}_g(c_1,c_2)^T, \nonumber \\
\mathcal{U}_s:&& (c_1,c_2^{\dagger})^T\to\mathcal{U}_s(c_1,c_2^{\dagger})^T.
\eea
It follows that states which are singlets under $\mathrm{SU}(2)_\text{gauge}$ are doublets under $\mathrm{SU}(2)_\text{spin}$ and vice versa. Specifically, there are two distinct kinds of vortices, labeled by the fermion parity $(-1)^{c_1^{\dagger}c_1+c_2^{\dagger}c_2}$: both have $\tilde{\T}^2=-1$, but one transforms as $(0,1/2)$ under $\mathrm{SU}(2)_\text{gauge} \times \mathrm{SU}(2)_\text{spin}$, and the other as $(1/2,0)$.

We now restore time-reversal symmetry by condensing double-vortices  that transform trivially under both $\mathrm{SU}(2)_\text{gauge} \times \mathrm{SU}(2)_\text{spin}$ and $\tilde \T$, giving $\mathbb{Z}_2$ topological order on the surface \cite{z2long,bfn}. Single vortices with even and odd fermion parity yield mutual semions which we denote $e$ and $\tilde{m}$ respectively. Both are Kramers bosons ($\T^2=-1$), and $e$ transforms as $(0,1/2)$ under $\mathrm{SU}(2)_\text{gauge} \times \mathrm{SU}(2)_\text{spin}$ while $\tilde{m}$ transforms as $(1/2,0)$.  Their bound state, $\tilde \epsilon$, is non--Kramers, fermionic, and transforms as $(1/2,1/2)$. 

So far, our treatment of the surface has neglected the confining gauge field.\footnote{The fermions in the bulk are confined, giving a non-fractionalized bulk state. It is known \cite{srl} that the $\mathrm{SU}(2)$ gauge theory has a $\theta$-term at $\theta=(\nu'/2)\pi$. For $\nu'=4$, as here, we have $\theta=2\pi$ which has the same physics as at $\theta = 0$.  The confined state then can preserve time reversal symmetry.  In contrast if $\nu' = 2$, we will have $\theta = \pi$ and the resulting confined phase of the $SU(2)$ gauge theory must be   
non-trivial in some way. It either breaks time reversal or becomes a quantum spin liquid with long range entanglement.}  When we take it into account, only excitations that are neutral under $\mathrm{SU}(2)_\text{gauge}$ survive. In addition to  $e$, these include bound states $m=\psi \tilde \epsilon$ and $\epsilon=\psi \tilde m$ got by attaching a $\psi$ fermion to $\tilde m$ and $\tilde\epsilon$. This shifts the self-statistics, so $m$ is bosonic while $\epsilon$ is fermionic (all three particles are mutual semions). Since $\epsilon$ is the bound state of $e$ and $m$ (and its properties follow from this) we do not discuss it further. Note that $m=\psi \tilde \epsilon$ is Kramers since $\psi$ is.

The upshot is that the surface topological order surviving after `gauge neutralization' has an $e$ particle that is Kramers and spin-doublet, and an $m$ particle that is Kramers but spin-singlet.  Since both $e$ and $m$ are Kramers bosons, this state is indeed the $eTmT$ phase, like the wavefunction discussed in Sec.~\ref{loop gas section}.

However if spin-rotation symmetry is preserved, a finer classification is possible, under which the present state is dubbed $eCTmT$, where the `$C$' indicates that $e$ is a spin doublet \cite{hmodl}.\footnote{The $eCTmT$ state is topologically equivalent to the combination of a generic $eTmT$ state and the state $eCmT$; the analogue of the latter for $\mathrm{U}(1)$ spin symmetry is discussed in Sec.~\ref{eCmT}.} This finer classification emphasises a difference between the $eTmT$ state constructed here, in which $e$ is a spin doublet and $m$ is not, and that constructed in Sec.~\ref{loop gas section}, where both $e$ and $m$ are spin doublets. 

Like the 1D example of the previous section, the  cubic lattice construction violates inversion symmetry at the free fermion level (this is inevitable if $\nu$ is nontrivial \cite{Lu Lee Inversion}) but the resulting spin state is inversion symmetric as a result of gauge invariance. Here, the hopping term in $H_\text{MF}$ is invariant under inversion, while for an appropriate choice of $\chi'$ the pairing terms change sign under inversion. Therefore inversion can be restored by combining it with the gauge transformation $f\rightarrow i f$, i.e. $\eta\rightarrow i \rho^y \eta$. (With the arrow conventions of Sec.~\ref{Further details of AKLT state}, the fluctuating AKLT state is also inversion symmetric.)

\subsection{Diamond lattice}

Next we consider parton theories on the diamond lattice, making use of the band structure of  Ref.~\onlinecite{srl}. First  we construct a theory with an $\mathrm{SU}(2)$ gauge field which naturally confines (Sec.~\ref{eCmT}).  The resulting state is a topological paramagnet which requires both time-reversal and $XY$-spin rotation symmetry to be protected. Then in Sec.~\ref{eCTmT} we construct a $\mathrm{U}(1)$ gauge theory, which confines at strong coupling. The confined state is a topological paramagnet which only requires time-reversal symmetry.

\subsubsection{Topological $XY$ paramagnet from ${SU}(2)$ gauge theory}
\label{eCmT}
 The mean field Hamiltonian is
\be
\label{diaband}
H_\text{MF}=\sum_{\langle ij\rangle}t\eta_i^{T}\rho^y\eta_j+\sum_{\langle\langle ij\rangle\rangle}t'_{ij}\eta_i^T\rho^y\eta_j+\sum_{\langle\langle ij \rangle\rangle}\Delta_{ij}\eta^T_i\rho^x\tau^y\eta_j,
\ee
 where the nearest-neighbor hopping $t$ is isotropic, while the next-nearest-neighbor hopping $t'_{ij}$ and pairing $\Delta_{ij}$ follow the patterns discussed in Ref.~\onlinecite{srl}. Notice that the pairing term is a singlet in orbital space, but is a triplet in spin space. Hence the spin rotation symmetry is reduced from $\mathrm{SO}(3)$ down to $\mathrm{O}(2)$ rotations about the $S_y$ axis,  corresponding to XY anisotropy in the spin model.

We again calculate the nontrivial Wison loops: the simplest nontrivial ones consist of three links and are proportional to $\rho^y$ and $\rho^x\tau^y$. The unbroken gauge group is generated by
\be
\Gamma_\text{diamond}=\{ \rho^y\tau^x,\tau^y,\rho^y\tau^z \}.
\ee
These are precisely the orbital $\mathrm{SU}(2)$ generators $\vec\Omega$.

One can  implement time-reversal symmetry $\T$ as $\eta\to \rho^z\sigma^y\tau^y\eta$, under which $\eta$ is non-Kramers ($\T^2=1$) and $\vec{S}\to-\vec{S}$ and of course $H_\text{MF}\to H_\text{MF}$.

As above we must check that the $\SP(4)$ gauge charges are all zero on average: $\langle\Gamma_i\rangle=0$. The unbroken gauge symmetry $\Gamma_\text{diamond}$ guarantees that $\langle\Gamma_i\rangle=0$ for all $i$ except for $\Gamma_1=\rho^y$, which is nothing but the total fermion occupation number (minus two). Fortunately the mean field Hamiltonian Eq.~\eqref{diaband} has a special lattice symmetry \cite{notesym} that sets $\langle\Gamma_1\rangle=0$. 

To determine the topology of the mean field band structure, it is convenient to consider the modified time-reversal symmetry $\mathcal{T}': \eta\to i\rho^z\tau^y\eta$ (with $\T'^2=-1$), which is the combination of time-reversal and spin rotation $i\sigma_y$. Fermions with different physical spins ($\eta_{\up}$ and $\eta_{\down}$) do not mix under the modified time-reversal. Furthermore, they are decoupled in the mean field Hamiltionian $H_\text{MF}$ and form two copies of an identical band. Therefore the topological index $\nu'$ is defined for each band separately. Now each band is identical to that studied in Ref.~\onlinecite{srl}, with $\nu'=4$. The total band therefore has $\nu'=8$, with four Dirac cones in total on the surface.

We now consider the surface Dirac theory with~$\mathrm{SU}(2)_\text{gauge}\times \mathrm{U}(1)_\text{spin}\times\mathcal{T}$ symmetry, with four Dirac cones in total:

\be
H=\psi^{\dagger}(p_x\mu_x+p_y\mu_z)\otimes\tau_0\otimes\sigma_0\psi,
\ee
with modified time-reversal
\be
\mathcal{T}':\psi\to i\mu_y\otimes\tau_0\otimes\sigma_0\psi^{\dagger},
\ee
gauge $\mathrm{SU}(2)$
\be
\mathcal{U}_g: \psi\to\mu_0\otimes \mathcal{U}_g\otimes\sigma_0\psi,
\ee
and spin $\mathrm{U}(1)$
\be
U_s(\theta): \psi\to e^{i\theta}\psi.
\ee
The actual time-reversal is $\T=U_s(\pi/2)\T'$:
\be
\mathcal{T}:\psi\to \mu_y\otimes\tau_0\otimes\sigma_0\psi^{\dagger}.
\ee

Now consider the gap term
\be
H_{\Delta}=i\Delta\,\psi\mu_y\otimes\tau_y\otimes\sigma_y\psi+h.c.,
\ee
which preserves both $\mathrm{SU}(2)_\text{gauge}$ and $\T$, but breaks $\mathrm{U}(1)_\text{spin}$. To restore the $\mathrm{U}(1)_\text{spin}$ symmetry and preserve the gap, we need to proliferate vortices in the order parameter field $\Delta$. It was shown in Ref.~\onlinecite{wpssc14,3dfSPT2,maxvortex} that the fundamental vortices have $\T^2=-1$, so condensing double-vortices gives a $\mathbb{Z}_2$ gauge theory, with $e$ being Kramers, $\tilde{m}$ being Kramers and $\mathrm{SU}(2)_\text{gauge}$-doublet, and $\tilde{\epsilon}$ being non-Kramers and $\mathrm{SU}(2)_\text{gauge}$-doublet. We can then gauge-neutralize the particles by binding $\psi$ fermions to $\tilde{m}$ and $\tilde{\epsilon}$. The neutralized theory then has $e$ being Kramers, and $m=\tilde{\epsilon}\psi$ being non-Kramers (recall that $\T^2=1$ on $\psi$) but carrying spin-$1/2$ under $\mathrm{U}(1)_\text{spin}$ due to the $S_y$ spin carried by $\psi$. This state is dubbed $eCmT$ in Ref.~\onlinecite{hmodl}. 

The fermions will be confined once the fluctuation of the $\mathrm{SU}(2)$ gauge field is introduced, and we obtain a non-fractionalized bulk state. On the surface, the $eCmT$ topological order survives the confinement, since all the non-trivial quasi-particles in the theory are gauge-neutral and are hence decoupled from the gauge field. We have thus obtained the $eCmT$ topological paramagnet. 

As a side note, if the spin-$1$ operators are pseudo-spins such that $\T: \{S_x,S_y,S_z\}\to\{S_x,-S_y,S_z\}$, then the modified time-reversal $\mathcal{T}': \eta\to i\rho^z\tau^y\eta$ (with $\T'^2=-1$) could represent the physical time-reversal symmetry. In this case we obtain a topological paramagnet that requires time-reversal only, as will be shown in Sec \ref{eCTmT}.

\subsubsection{Stable $U(1)$ quantum spin liquids and topological paramagnets}
\label{eCTmT}
The parton construction of course also gives access to stable quantum spin liquid phases. Of particular interest to us is a time reversal symmetric $\mathrm{U}(1)$ quantum spin liquid phase on the diamond lattice. For greater generality we allow for full $\mathrm{SU}(2)$ spin symmetry.  As usual such a phase has a gapless emergent photon. In addition it has a gapped fermionic spin-$1/2$ Kramers doublet spinon which has internal `electric' charge\footnote{This `electric' charge couples to the emergent photon in this spin liquid, and not to physical external electromagnetic fields.}    and a gapped bosonic spin-$0$ magnetic monopole that transforms to an antimonopole under time reversal.   We will give the spinons the band structure of a topological superconductor (as in previous sections).  The resulting quantum spin liquid phase then inherits the non-trivial surface states of the topological superconductor. The relevance to the present paper comes from asking about the confined phase that results when the 
magnetic monopole is 
condensed. We show below that this is the $eCTmT$ topological paramagnet. 

SPT phases in 3D have been accessed previously through confinement of emergent $\mathrm{U}(1)$ gauge fields \cite{3dpartonu1}. However in these previous studies the confinement was  achieved in a highly non-trivial way  involving the condensation of dyons (bound states of magnetic and electric charges). The novel aspect of our construction is that the confinement is achieved directly by  simply condensing the magnetic monopole, which will result from the usual dynamics of the gauge field at strong coupling.

Consider the following mean field ansatz:
\bea
\label{diaband2}
H_\text{MF}=&&\sum_{\langle ij\rangle}t\eta_i^{T}\rho^y\eta_j+\sum_{\langle\langle ij\rangle\rangle}t'_{ij}\eta_i^T\rho^y\eta_j+\sum_{\langle\langle ij \rangle\rangle}\Delta_{ij}\eta^T_i\rho^x\sigma^y\eta_j  \nonumber \\
&&+\sum_i\Delta'\eta^T_i\rho^x\sigma^y\eta_i+\sum_{\<ij\>,\, i\in A}it''\eta_i^T\rho^y\tau^y\eta_j,
\eea
where the nearest-neighbor hopping $t$ and on-site pairing $\Delta'$ are uniform and isotropic, while the next-nearest-neighbor hopping $t'_{ij}$ and pairing $\Delta_{ij}$ follow the patterns discussed in Ref.~\onlinecite{srl}. Note that the first two terms are the same as in Eq.~\eqref{diaband}, and the third is got by exchanging the role of orbital and physical spin. Contrary to Eq.~\eqref{diaband}, the pairing term $\Delta$ is a singlet in physical spin and a triplet in orbital space, so the full spin rotation symmetry is preserved. The nearest-neighbor antisymmetric hopping term $t''$ is introduced to reduce the gauge symmetry, and does not affect the other arguments in this section as long as it is kept small.
 
The simplest nontrivial Wilson loops are proportional to $\rho^y$, $\rho^x\sigma^y$ and $\rho^y\tau^y$. The resulting unbroken gauge group is a $\mathrm{U}(1)$ generated by $\tau^y$.

We implement time-reversal symmetry $\T$ through $\eta\to i\rho^z\sigma^y\eta$ (which has $\T^2=-1$). It is straightforward to check that $\vec{S}\to-\vec{S}$ and $H_\text{MF}\to H_\text{MF}$ under the chosen time-reversal symmetry. Moreover, the $\mathrm{U}(1)$ gauge charge $\tau^y$ is also odd under $\T$, which allows  for topologically non-trivial band structures for the partons.

We now check that $\langle\Gamma_i\rangle=0$. 
The unbroken $\mathrm{U}(1)$ gauge symmetry and time-reversal guarantee that $\langle\Gamma_i\rangle=0$ for all $i$ except for $\rho^y$ and $\rho^x\sigma^y$, which are nothing but the total fermion occupation number (minus two) and the real part of the on-site pairing. The lattice symmetry \cite{notesym} again sets $\langle\rho^y\rangle=0$. 
For the on-site pairing amplitude, there is no symmetry to set it to zero automatically. We must therefore adjust the on-site pairing term $\Delta'$ in Eq.~\eqref{diaband2} to make it zero on average \cite{onsite pairing}.

To determine the topology of the mean field band structure, notice that fermions with different orbital indices ($\tau$ indices) do not mix under time-reversal $\T: \eta\to i\rho^z\sigma^y\eta$. They are also decoupled in the mean field Hamiltionian $H_\text{MF}$, forming two copies of an identical band. Therefore the topological index $\nu'$ is defined for each band separately. Now each band is almost identical to that studied in Ref.~\onlinecite{srl}, with $\nu'=4$. The total band therefore has $\nu'=8$, with four Dirac cones in total on the surface.

We now consider fluctuations of the $\mathrm{U}(1)$ gauge field. In the weak coupling regime the gauge theory is deconfined, and we have a stable $\mathrm{U}(1)$ quantum spin liquid phase. 
The spinon band structure has time reversal protected surface states that provide a distinction between this spin liquid and more conventional ones. 
For a compact $\mathrm{U}(1)$ gauge theory, there are always gapped magnetic monopole excitations in the theory.  In Ref.~\onlinecite{wpssc14,3dfSPT2} it was shown that for the spinon band structure we have here,  this magnetic monopole is a spin-$0$ boson that simply transforms into an antimonopole under time reversal. 

As the gauge coupling strength increases, the monopole mass gap decreases and eventually becomes zero. The monopoles will then condense and confine the gauge theory. 
 The trivial symmetry properties of  the monopole implies that this condensate does not break $\T$ or the physical spin $\mathrm{SU}(2)$ (if present). The confined state is thus a non-fractionalized symmetry preserving paramagnet. To determine which SPT phase the paramagnet belongs to, we need to examine the surface state in more detail. The argument is largely parallel to that in Sec.~\ref{cube}, with the simple modification that the $\mathrm{SU}(2)$ gauge symmetry discussed in Sec.~\ref{cube} is reduced to $\mathrm{U}(1)$. The conclusion remains the same: the paramagnet is the nontrivial SPT dubbed $eCTmT$ in Ref.~\onlinecite{hmodl}. 
The representative surface state is a gapped $\mathbb{Z}_2$ topological order, with $e$ being Kramers and spin-doublet, and $m$ Kramers but spin-singlet. (If the spin-rotation symmetry is broken, this becomes a generic $eTmT$ state.)

\subsection{Spin wavefunctions}
\label{spin wavefunction section}

The parton constructions suggest spin wave functions that may be useful as variational states in future work on specific microscopic models.  Following the standard 
procedure \cite{Wenbook} we construct a spin wave function 
  from the mean-field fermion wave function $|\Psi_{\text{MF}}\rangle$ by projecting onto the subspace obeying the constraints $\sum_{a\alpha} f^{\dagger}_{a\alpha}f_{a\alpha}=2$ and $\sum_{ab\alpha}f^{\dagger}_{a\alpha}\vec \tau_{ab}f_{b\alpha}=0$:
\be
|\Psi_{\text{spin}}\rangle=\mathcal{P} |\Psi_{\text{MF}}\rangle.
\ee
Such a projection is expected to roughly mimic the effect of gauge fluctuations. For the states constructed in Sec.~\ref{cube} and \ref{eCmT}, the $\mathrm{SU}(2)$ gauge fluctuations  automatically confine the states. We therefore expect the projected wave functions to represent the confined spin SPT states. 
For the state in Sec.~\ref{eCTmT}, the $\mathrm{U}(1)$ gauge field is deconfined at weak coupling, and confines to an SPT state at strong coupling. 
So it is not clear a priori whether the projected wave function will give the $U(1)$ quantum spin liquid state or the confined SPT state.

These spin wave functions are alternate possibilities to the loop gas wave functions described in the first part of the paper. While the loop gas wave functions are physically appealing they are likely not very tractable numerically due to the linking signs. The parton wave functions, on the other hand, may be studied through variational Monte Carlo calculations though the physical connection to SPT physics is less directly obvious.  This situation is similar to existing descriptions of quantum spin liquid phases through either loop gases (string-nets) or through partons which each have their advantages and disadvantages. 

For the topological paramagnets, at present we do not have a direct connection between the parton and loop gas wavefunctions. Establishing such a connection is a target for future work, and will confirm the general correctness of the projected wave functions as faithfully capturing the state accessed through the parton description.

\section{Discussion: Towards models and materials}

We have emphasized that frustrated spin-$1$ magnets in 3D may be fruitful in the search for spin SPT phases.  

In the ongoing search for quantum paramagnetism in frustrated systems, the bulk of the attention has focused on spin-$1/2$ systems.  This is guided by the intuition that increasing the spin only leads to more `classical' physics and hence to a greater tendency to order. Caution however is required in taking this intuition too seriously. 
In one dimension the spin-$1/2$ chain is almost antiferromagnetically ordered (power law correlations) while the spin-$1$ chain is a good paramagnet with a spin gap.  This has the following amusing  consequence. Consider a two--dimensional rectangular lattice with nearest neighbor antiferromagnetic interactions:
\begin{equation}
H_\text{rect} = J_\parallel \sum_{{\bf  r} } \vec S_{\bf r} \cdot \vec S_{{\bf r} + {\bf x}} + J_\perp \sum_{{\bf  r} } \vec S_{\bf r} \cdot \vec S_{{\bf r}+ {\bf y}}
\end{equation}
For $J_\parallel = J_\perp$ the model is antiferromagnetically ordered for all spin $S$. When $\frac{J_\perp}{J_\parallel}$ is decreased from $1$ the spin-$1/2$ model stays ordered 
unless $J_\perp = 0$.  The spin-$1$ model on the other hand becomes a spin gapped paramagnet below a  non-zero critical value of $\frac{J_\perp}{J_\parallel}$. So there is a range of parameters in this 2D model where the spin-$1$ system is a quantum paramagnet although the spin-$1/2$ system has long range Neel order.

There are some interesting examples of  frustrated spin-$1$ magnets  --- most notably $\mathrm{NiGa}_2\mathrm{S}_4$ and $\mathrm{Ba}_3\mathrm{NiSb}_2\mathrm{O}_9$, in both of which the spin-$1$ $\mathrm{Ni}$ ion forms a triangular lattice \cite{nakatsuji, cheng li balicas}.  Apart from new and interesting kinds of quantum spin liquids, spin-$1$ magnets may also harbor novel broken symmetry states (such as spin nematics \cite{NiGa2S4 nematic papers}) more naturally than their spin-$1/2$ counterparts.  To this we add the SPT phase discussed in this paper as a possible fate for a frustrated 3D spin-$1$ magnet.

Our results suggest a route to guessing possible microscopic models that might harbor an SPT phase.  Starting from the parton mean field Hamiltonian we can write down a lattice gauge theory that captures fluctuations. A strong coupling expansion of this lattice gauge theory will result in a spin Hamiltonian  which may then be in the same phase as the same lattice gauge theory at weaker coupling. Such an approach has previously been successfully used to write down lattice models for various spin liquid phases. Given that we are interested here in confined phases we may be cautiously optimistic that a similar approach has an even better chance of resulting in spin models for the SPT phases. As an application let us consider the diamond lattice parton construction. With full $\mathrm{SU}(2)$ spin symmetry, the mean field state of Section \ref{eCTmT} suggests (at leading order of the strong coupling expansion in the resulting $\mathrm{U}(1)$ gauge theory) an 
interesting frustrated spin-1 model:  the ``$J_1$--$J_2$" antiferromagnet on the diamond lattice\footnote{Strictly speaking, the $J_2$-coupling obtained from the previous mean-field ansatz should be anisotropic. It is not clear whether this anisotropy is in reality essential  for realizing the topological paramagnet.}: 
\begin{equation}
H = J_1\sum_{\langle r r'\rangle} \vec S_r \cdot \vec S_{r'} + J_2 \sum_{\langle \langle r r'\rangle \rangle} \vec S_r \cdot \vec S_{r'} 
\end{equation}
The next-nearest neighbour coupling $J_2$ introduces frustration.  Indeed classically once $J_2 > \frac{J_1}{8}$ there are an infinite number of degenerate ground states \cite{bergmann} that are not related by global spin rotation. For large spin, it has been argued that the ground state is magnetically ordered as a result of quantum order by disorder \cite{bernier}. The ground state for $S = 1$ (or $S = 1/2$) is not known. The SPT paramagnet discussed in this paper is a candidate. The various descriptions we have provided should be a  useful guide in future numerical studies should a paramagnetic ground state be found for this model. 

It is interesting to note that since the diamond lattice is 4--fold coordinated classical 2-sublattice Neel order is likely to be more easily destabilized by frustration/quantum fluctuations than in the cubic lattice.  Thus the $J_1$--$J_2$ diamond magnet for low spin ($S = 1/2$ or $1$) may be an excellent candidate to find an interesting quantum paramagnetic ground state. 

The frustrated diamond lattice model appears to describe well \cite{bergmann} the physics of the spinel oxide materials $\mathrm{MnAl_2O_4}$ and $\mathrm{CoAl_2O_4}$ \cite{tristan cubic spinels} which belong to a general family of materials of the form $\mathrm{AB_2O_4}$. The $A$ site forms the diamond lattice and is magnetic. The $\mathrm{Mn}$ and $\mathrm{Co}$ compounds have $S = \frac{5}{2}$ and $S = \frac{3}{2}$ respectively. In searching for a material that realizes the $S = 1$ model it is natural then to consider $\mathrm{NiAl_2O_4}$. Here $\mathrm{Ni}$ is expected to be in a $d^8$ $\mathrm{Ni}^{2+}$ configuration and have spin-$1$. However the $A$ site is tetrahedrally coordinated, and in the resulting crystal field, the $\mathrm{Ni}^{2+}$ ion will have orbital degeneracy in addition to spin-$1$. Further spin-orbit coupling will split the resulting spin-orbital Hilbert space and the physics of the lattice will be determined by its competition with inter-site spin/orbital exchange. Thus $\mathrm{NiAl_2O_4}$ will not be simply 
described by a spin-$1$ diamond lattice model. Work  toward obtaining an appropriate `spin-orbital' model for $\mathrm{NiAl_2O_4}$ is currently in progress \cite{seigenfeld}. It remains to be seen whether the presence of the orbital degrees of freedom aids or hinders the formation of paramagnetic states. 

In any case we hope that these considerations motivate an experimental search for and study of frustrated spin-$1$ magnets.

\section{Acknowledgements}  AN thanks Jeongwan Haah and Curt von Keyserlingk for useful discussions and acknowledges the support of a fellowship from the Moore Foundation. TS and CW acknowledge support from NSF DMR-130574.  TS was partially supported by a Simons Investigator award from the Simons Foundation.



\begin{thebibliography}{200}

\bibitem{sllee} P. A. Lee, Science 321, 1306 (2008).
\bibitem{slbalents} L. Balents, Nature 464, 199 (2010).
\bibitem{Wenbook} X. G. Wen, Quantum Field Theory Of Many-Body Systems, Oxford University Press (2004).
\bibitem{avts12} A. Vishwanath, T. Senthil,  Phys. Rev. X {\bf 3}, 011016 (2013). 
 \bibitem{1dsptclass} F. Pollmann, A. M. Turner, E. Berg, and M. Oshikawa, Phys. Rev. B {\bf 81}, 064439 (2010); A. M. Turner, F. Pollmann, and E. Berg, Phys. Rev. B {\bf 83},
075102 (2011);  X. Chen, Z.-C. Gu, and X.-G. Wen, Phys. Rev. B {\bf 83},
035107 (2011); N. Schuch,  D. P\'erez-Garcia, and I. Cirac, Phys. Rev. B {\bf 84}, 165139, (2011).
\bibitem{1dz8} L. Fidkowski and A. Kitaev, Phys. Rev. B {\bf 83}, 075103 (2011)
\bibitem{chencoho2011} X. Chen, Z.-C. Gu, Z.-X. Liu, X.-G. Wen, Science {\bf 338}, 1604 (2012); Phys. Rev. B {\bf 87}, 155114 (2013).
\bibitem{atav13} A. M. Turner and A. Vishwanath, arXiv:1301.0330 (2013).
\bibitem{sptannrev} T. Senthil,  arXiv:1405.4015 (2014).
\bibitem{hmodl} C. Wang and T. Senthil,  Phys. Rev. B {\bf 87}, 235122 (2013).
\bibitem{burnellbc} F. J. Burnell, X. Chen, L. Fidkowski, A. Vishwanath,  arXiv:1302.7072 (2013). 
\bibitem{rosch} F. Anfuso and A. Rosch, Phys. Rev. B {\bf 75}, 144420 (2007).
 \bibitem{wpssc14} C. Wang, A. C. Potter, and T. Senthil, Science {\bf 343}, 6171 (2014).
 \bibitem{kitaev toric code}   A. Y. Kitaev, Ann. Phys. {\bf 303}, 2 (2003).
 \bibitem{lesiknote} The same model has also been independently studied by S. Geraedts and O. Motrunich (private communication). 
 \bibitem{cenkeetal} C. Xu, F. Wang, Y. Qi, L. Balents and M. P. A. Fisher, Phys. Rev. Lett. {\bf 108}, 087204 (2012).
 \bibitem{AKLT paper} I. Affleck, T. Kennedy, E. H. Lieb, and H. Tasaki, Phys.
Rev. Lett. {\bf 59}, 799 (1987); Commun. Math. Phys. {\bf 115}, 477 (1988).
 \bibitem{kapustin thorngren higher symmetry}  A. Kapustin and R. Thorngren, arXiv: 1309.4721 (2013).
 \bibitem{burnell et al phase transitions} F. J. Burnell, C. W. von Keyserlingk, and S. H. Simon, Phys. Rev. B {\bf  88} 235120 (2013).
 \bibitem{walker wang} K. Walker and Z. Wang, Front. Phys. {\bf 7}, 150 (2012).
\bibitem{keyserlingk et al surface anyons} C. W. von Keyserlingk, F. J. Burnell, and S. H. Simon, Phys. Rev. B {\bf 87}, 045107 (2013).
 \bibitem{chenanomaloussymm} X. Chen, F. Burnell, A. Vishwanath, and L. Fidkowski, arXiv:1403.6491 (2014).
 \bibitem{Yao Fu Qi} H. Yao, L. Fu, and X.-L. Qi, arXiv:1012.4470 (2010).
\bibitem{Huang et al SET} C. Y. Huang, X. Chen, and F. Pollmann, Phys. Rev. B {\bf 90}, 045142 (2014).
\bibitem{Li et al RAKLT} W. Li, S. Yang, M. Cheng, Z.-X. Liu, and H.-H. Tu, Phys. Rev. B {\bf 89}, 174411 (2014).
  \bibitem{avdomain} X. Chen, Y.-M. Lu, A. Vishwanath,  Nature Comm. {\bf 5}, 3507 (2014). 
 \bibitem{wang xu two orbital boson} F. Wang and C. Xu, arXiv:1110.4091 (2011).
\bibitem{fradkin book} E. Fradkin, Field Theories of Condensed Matter Physics, 2nd edition, Cambridge University Press (2013).
\bibitem{xuts13} C. Xu and T. Senthil, Phys. Rev. B {\bf 87}, 174412 (2013). 
 \bibitem{cubic RVB numerics} A. Fabricio Albuquerque, F. Alet, and R. Moessner, Phys. Rev. Lett. {\bf 109}, 147204 (2012).
\bibitem{reduced density matrix note} In the `pure loop' state of Sec.~\ref{pure loop section}, the reduced density matrix for a single sublattice is diagonal: $\rho_A\propto \sum_{\CC_A} \ket{\CC_A}\bra{\CC_A}$. The reduced density matrix for the AKLT based state does not have a simple form, but by analogy we expect a suppression of off-diagonal elements. (In the artificial limit where they are completely suppressed, the spins are trivially short-range correlated.)
 \bibitem{auerbach} A. Auerbach, Interacting Electrons and Quantum Magnetism, Springer (1994).
\bibitem{2dparton} P. Ye and X.-G. Wen, Phys. Rev. B {\bf 87}, 195128 (2013);
Y.-M. Lu and D.-H. Lee, arXiv:1210.0909 (2012); T. Grover and A. Vishwanath, Phys. Rev. B {\bf 87}, 045129 (2013); 
\bibitem{cenkeprtn2d} J. Oon, G. Y. Cho, and C. Xu, arXiv:1212.1726 (2012).
\bibitem{3dpartonu1} P. Ye and X.-G. Wen, Phys. Rev. B {\bf 89}, 045127 (2014); M. A. Metlitski, C. L. Kane and M. P. A. Fisher, unpublished.
\bibitem{3dpartonz2} Z. Bi, A. Rasmussen, Y. You, M. Cheng and C. Xu, arXiv:1404.6256 (2014).
 \bibitem{tenfold} A. Kitaev, arXiv:0901.2686 (2009); S. Ryu, A. P. Schnyder, A. Furusaki and A. W. W. Ludwig, New J. Phys. {\bf 12}, 065010 (2010).
\bibitem{fidkowski3d} L. Fidkowski, X. Chen and A. Vishwanath, Phys. Rev. X {\bf 3}, 041016 (2013).
\bibitem{3dfSPT2} C. Wang and T. Senthil, Phys. Rev. B {\bf 89}, 195124 (2014).
\bibitem{maxvortex} M. A. Metlitski, L. Fidkowski, X. Chen and A. Vishwanath, arXiv:1406.3032 (2014).
\bibitem{hrv} P. Hosur, S. Ryu and A. Vishwanath, Phys. Rev. B, {\bf 81}, 045120 (2010).
\bibitem{srl} A. P. Schnyder, S. Ryu and A.  W. W. Ludwig, Phys. Rev. Lett. {\bf 102}, 196804 (2009).
\bibitem{1dparton} Z.-X. Liu, Y. Zhou, H.-H. Tu, X.-G. Wen and T.-K. Ng, Phys. Rev. B {\bf 85}, 195144 (2012).
\bibitem{Kitaev chain} A. Y. Kitaev, Phys. Usp. {\bf 44}, 131 (2001).
\bibitem{tangwen} E. Tang and X.-G. Wen, Phys. Rev. Lett. {\bf 109}, 096403 (2012).
\bibitem{Greiter Schnells Thomale} M. Greiter, V. Schnells, and R. Thomale, arXiv:1402.5262 (2014).
\bibitem{hughes} M. F. Lapa, J. C. Y. Teo and T. L. Hughes, arXiv:1409.1234 (2014).
 \bibitem{z2long} T. Senthil and M. P. A. Fisher, Phys. Rev. B {\bf 62}, 7850 (2000).
\bibitem{bfn} L. Balents, M. P. A. Fisher, and C. Nayak, Phys. Rev. B {\bf 60}, 1654 (1999). 
\bibitem{Lu Lee Inversion} Y.-M. Lu and D.-H. Lee, arXiv:1403.5558 (2014).
\bibitem{notesym}The symmetry is the combination of a $\pi/2$ rotation along $\hat{z}$, a reflection $z\to-z$, then followed by a particle-hole transformation $c_{\alpha,\delta,a}=\sigma^z_{\alpha\beta}\tau^z_{\delta\lambda}\gamma^z_{ab}c^{\dagger}_{\beta,\lambda,b}$, where $\{\alpha,\beta\}$ denote the spin, $\{\delta,\lambda\}$ denote the orbital, and $\{a,b\}$ denote the sub-lattice index ($\gamma^z$ is a Pauli matrix acting on the sub-lattice indicies).
\bibitem{onsite pairing}  We must check that this does not close the gap. The total pairing term in momentum space $\Delta_{\vec{k}}+\Delta_{\vec{k}}'$ must not be positive (or negative) definite, since the onsite pairing vanishes: $\langle f_{i\uparrow}f_{i\downarrow}\rangle=0$. It is easy to show that this requires $|\Delta'|<12|\Delta|$. One can then show that such a value of $\Delta'$ can never close the gap opened by $\Delta$. Therefore the mean field Hamiltionian Eq.~\eqref{diaband2} can be smoothly connected to a Hamiltionian with no $\Delta'$ term without closing the gap.
\bibitem{nakatsuji} S. Nakatsuji et al., Science {\bf 309}, 1697 (2005).
\bibitem{cheng li balicas} J. G. Cheng et al., Phys. Rev. Lett. {\bf 107} 197204 (2011).
\bibitem{NiGa2S4 nematic papers} H. Tsunetsugu and M. Arikawa, J. Phys. Soc. Jpn. {\bf 75},
083701 (2006); A. Lauchli, F. Mila and K. Penc, Phys. Rev. Lett. {\bf 97}, 087205 (2006); S. Bhattacharjee, V. B. Shenoy, and T. Senthil, Phys. Rev. B {\bf 74}, 092406 (2006); E. M. Stoudenmire, S. Trebst and L. Balents, Phys. Rev. B {\bf 79}, 214436 (2009).
\bibitem{bergmann} D. Bergman, J. Alicea, E. Gull, S. Trebst, L. Balents, Nature Physics {\bf 3}, 487 (2007).
\bibitem{bernier} J.-S. Bernier, M. J. Lawler and Y. B. Kim, Phys. Rev. Lett. {\bf 101}, 047201 (2008).
\bibitem{tristan cubic spinels} N. Tristan et al., Phys Rev. B {\bf 72}, 174404 (2005).
\bibitem{seigenfeld} A. Seigenfeld and T. Senthil (in progress). 
\end{thebibliography}
\end{document}